\documentclass[12pt]{iopart}
\usepackage{graphicx}
\usepackage{epsfig}

\begin{document}

\title[Perspectives of the disproportionation driven superconductivity]{Perspectives of the disproportionation driven superconductivity in strongly correlated 3d compounds}

\author{A.S. Moskvin}

\address{Department of Theoretical Physics, Ural Federal University, 620083 Ekaterinburg,  Russia}
\ead{alexandr.moskvin@usu.ru}
\begin{abstract}
Disproportionation in 3d compounds can give rise to an unconventional electron-hole Bose liquid with a very rich phase diagram from a Bose metal, charge ordering insulator to an inhomogeneous Bose superfluid.
Optimal conditions for the  disproportionation driven high-T$_c$ superconductivity  are shown to realize  only for several Jahn-Teller $d^n$ configurations that permit the formation of well defined local composite bosons. These are the high-spin $d^4$, low-spin $d^7$, and $d^9$ configurations given the octahedral crystal field, and the $d^1$, high-spin $d^6$ configurations given the tetrahedral crystal field. The  disproportionation reaction has a peculiar "anti-Jahn-Teller" character lifting the bare orbital degeneracy. Superconductivity in the $d^4$ and $d^6$ systems at variance with $d^1$, $d^7$, and $d^9$ systems  implies an unavoidable coexistence of the spin-triplet composite bosons and a magnetic lattice. 
We argue that unconventional high-T$_c$ superconductivity observed in quasi-2D cuprates with tetragonally distorted CuO$_6$ octahedra and iron-based layered pnictides/chalcogenides with tetrahedrally coordinated Fe$^{2+}$ ions can be a key argument supporting the disproportionation scenario is at work in these compounds.
\end{abstract}
\maketitle



\section{Introduction}

The origin of high-T$_c$ superconductivity\,\cite{Muller} is presently still a matter of great
controversy. Copper oxides start out life as insulators in contrast with BCS
superconductors being conventional metals. Unconventional behavior of these materials under charge doping, in particular, a remarkable interplay of charge, lattice, orbital, and spin degrees of freedom, strongly differs from that of ordinary metals and merely resembles that of a doped semiconductor. Novel non-copper based layered high-T$_c$ materials such as iron oxyarsenide LaOFeAs\,\cite{Kamihara} reveal normal and superconducting state properties very different from that of standard electron-phonon coupled "conventional" \, superconductors.

We believe that unconventional behavior of high-T$_c$ superconductors can be consistently explained in frames of a so called dielectric scenario\,\cite{Moskvin,Moskvin-LTP,Moskvin-11} that implies an instability of the respective parent compound with regard to  \emph{d-d} charge transfer (CT) fluctuations with formation of local composite bosons. This implies realisation of the Ogg-Schafroth's real-space pairing model for the Bose-Einstein superconducting condensation\,\cite{Ogg,Schafroth}.
At first sight the \emph{d-d} disproportionation 
\begin{equation}
3d^n+3d^n \rightarrow 3d^{n+1}+3d^{n-1}	
\label{disp}
\end{equation}
whose energy is usually estimated to be $U$ ($U_{dd}$), mean intra-atomic electron-electron
 (Mott-Hubbard) repulsion energy, is rather uncommon and a priori much more expensive in 
  narrow-band 3d transition-metal compounds  as compared with compounds containing broad-band  $6s^1$ ions such as $Bi^{4+}$, $Pb^{3+}$ or $Tl^{2+}$ due to their strong tendency to form $6s^0$ and $6s^2$ closed shells\,\cite{negative-U}. Experimental values of the minimal energies  of the $d$\,-\,$d$ CT, or Mott-Hubbard transitions in different 3d oxides as usually derived from the optical gap measurements, $\Delta_{dd}^{opt}$\,$\sim$\,2-4\,eV (see, e.g., Ref.\,\cite{dd-CT} and references therein) yield effective $U$'s which are sizeably less than typical $U_{dd}$\,$\sim$\,5-10\,eV, however,  far from the so-called "negative-$U$" regime\,\cite{negative-U}. Yet $\Delta_{dd}^{opt}$ is a minimal energy cost of the optically excited $d$\,-\,$d$ disproportionation or electron-hole formation due to a direct Franck-Condon (FC) CT transition. The question arises, what is the energy cost for the thermal excitation of such a local \emph{d-d} disproportionation?  The answer implies first of all the knowledge of the relaxation energy, or  the energy gain due to the lattice polarization by the localized charges that can exceed 1\,eV\,\cite{Shluger}. 
For instance, in insulating cuprates such as La$_2$CuO$_4$  the true (thermal) $d$\,-\,$d$ charge transfer gap appears to be as small as 0.4-0.5\,eV rather than 1.5-2.0\,eV as derived from the optical gap measurements\,\cite{Moskvin-11}. In other words, parent cuprates should be addressed to be \emph{d-d} CT unstable systems.   

At present, the CT instability  with regard to the \emph{d-d} disproportionation and formation of  "negative-$U$" \, centers\,\cite{negative-U} is believed to be a rather
typical property for a number of perovskite 3d transition-metal oxides
 such as CaFeO$_3$, SrFeO$_3$, RNiO$_3$\,\cite{Mizokawa,Alonso}, furthermore,  in solid state chemistry  one consider tens of disproportionated systems\,\cite{Ionov}.
 
Classical disproportionation seems to be observed in the perovskite ferrate CaFeO$_3$, where the average formal valence of the iron ion is Fe$^{4+}$($3d^4$) is known to show a gradual charge disproportionation 2Fe$^{4+}$$\rightarrow$Fe$^{3+}$+Fe$^{5+}$   below 290\,K with antiferromagnetism below 115\,K according to M\"{o}ssbauer studies\,\cite{CaFeO3}. 
 The single magnetic hyperfine pattern for isostructural ferrate SrFeO$_3$ at 4\,K, on the other hand, indicates seemingly a rapid electron exchange between Fe$^{3+}$ and Fe$^{5+}$ ions, and the hyperfine field coincides approximately with the average value of the corresponding parameters for CaFeO$_3$. Probably, in this $3d^4$ oxide we meet with a manifestation of quantum  effects and formation of a quantum dimerized phase such as a "valence bond solid" composed of S- or P-type $neutral$ dimers described by  wave functions: 
 $$
\frac{1}{\sqrt{2}}\left[\Psi_1(3d^{n+1})\Psi_2(3d^{n-1})\pm \Psi_1(3d^{n-1})\Psi_2(3d^{n+1}\right]\, , 
$$
respectively. The $^{57}$Fe M\"{o}ssbauer measurements point to a charge disproportionation in 
Sr$_3$Fe$_2$O$_7$, which has a double-layered perovskite structure\,\cite{Sr3Fe2O7}.

Rare earth nickelates RNiO$_3$ (R is a trivalent rare earth ion) 
exhibit a first order  metal-insulator phase transition (MIT) upon cooling with distinct  signatures of the charge disproportionation\,\cite{Alonso,thermopower}. At T$_N$\,$\leq$\,T$_{MIT}$ these exhibit a further transition to a long-range antiferromagnetic  ordered state with a nonzero magnetic moment for one of the two nonequivalent Ni sites\,\cite{DyNiO3}.
A clear fingerprint of charge disproportionation is the breathing-type distortion of metal-oxygen octahedra since the different charge states of the transition-metal ion
take different ionic radii (see, e.g., Ref.\,\cite{Alonso}).

However, a large body of 3d materials does not manifest the charge instability so clearly as (Ca,Sr)FeO$_3$ or RNiO$_3$.
For instance,   the perovskite  manganites RMnO$_3$ with the same  as in ferrates $3d^4$ configuration of Mn$^{3+}$ ions reveal a "hidden" \, CT instability\,\cite{Good,Moskvin-09}. 
At first glance the disproportionation in manganese compounds is hardly possible since manganese atom does not manifest a "valence-skipping" \, phenomenon as, e.g. bismuth atom which can be found as Bi$^{3+}$ or Bi$^{5+}$, but not Bi$^{4+}$, with a generic bismuth oxide BaBiO$_3$ to be a well-known example of a charge disproportionated system (see, e.g., Ref.\,\cite{BaBiO3} and references therein). However, strictly speaking, sometimes manganese reveals a valence preference, e.g., while both Mn$^{2+}$ and  Mn$^{4+}$ are observed in MgO:Mn and CaO:Mn, the Mn$^{3+}$ center is missing\,\cite{MgO}.

The reason for valence skipping or valence preference observed for many elements still remains a mistery. Recently, Harrison\,\cite{Harrison} has argued that most likely traditional lattice relaxation effects, rather than any intra-atomic mechanisms\,\cite{negative-U} (specific behavior of ionization energies, stability of closed shells, strong screening of the high-charged states) are a driving force for disproportionation with formation of "negative-$U$" centers.

 Earlier it was argued that the disproportionated system can form an unconventional electron-hole Bose liquid whose phase diagram incorporates different phase states from classical, or chemical disproportionated state to quantum states, in particular, to the unconventional  Bose-superfluid (superconducting) state\,\cite{Moskvin-LTP}.
Regrettably physicists have paid remarkably little attention to the problem of valence disproportionation and negative-$U$ approaches ("chemical" route!) which are surely being grossly neglected in all present formal theoretical treatments of HTSC. 
Speaking about a close relation between disproportionation and superconductivity  it is worth noting the text-book example of BaBiO$_3$ system where we unexpectedly deal with the disproportionated Ba$^{3+}$+ Ba$^{5+}$ ground state instead of the conventional lattice of Ba$^{4+}$ cations\,\cite{negative-U}. The bismuthate can be converted to a superconductor by a nonisovalent substitution such as in Ba$^{2+}$$_{1-x}$K$^+$$_x$BiO$_3$. At present,  this system seems to be the only one where the unconventional superconductivity is related anyhow with the disproportionation reaction.

In the paper we concern ourselves  with great points 
of a so-called "disproportionation" scenario in 3d compounds which was addressed earlier by many authors, however, by now it was not properly developed as its consistent description in terms of conventional Hubbard type models is complicated, if any. 
Our final goal is to ascertain the criteria for  a motivated search of the 3d systems which are the most promising ones  as  high-temperature superconductors. 
 The paper is organized as follows. In Sec.II addressing a simple model of the CT unstable 2D system we demonstrate a potential of the disproportionation as a driving force for superconductivity. In Sec.III we analyze the features of the disproportionation reaction for different $3d^n$ pairs and do find $3d^n$ configurations optimal for the disproportionation driven superconductivity. In Sec.IV we address the electron structure of familiar 3d compounds with the optimal $3d^n$ configurations and point to high-T$_c$ superconductivity observed in quasi-2D cuprates and iron pnictides/chalcogenides to be a result of the \emph{d-d} disproportionation.  A short conclusion is made in Sec.V.

\section{Simple toy model of the mixed valence system and the disproportionation driven superconductivity}

\subsection{Pseudospin description of the model mixed-valent system}

Valent electronic states in  strongly correlated 3\emph{d} oxides manifest
both significant correlations and \emph{p}-\emph{d} covalency with a distinct trend to localisation of many-electron configurations formed by antibonding Me 3d-O 2p hybridized molecular orbitals. The localisation effects are particularly clear featured in the crystal field \emph{d}-\emph{d} transitions whose spectra just weakly vary from diluted to concentrated 3\emph{d} oxides. An optimal way to describe valent electronic states in  strongly correlated 3\emph{d} oxides is provided by quantum-chemical techniques such as the ligand field theory\,\cite{Ballhausen} which implies a crystal composed of a system of small 3d-cation-anion clusters. Naturally, such an approach has a number of 
shortcomings, nevertheless, this provides a clear physical
picture of the complex electronic structure and the energy
spectrum, as well as the possibility of a quantitative modelling.
In a certain sense the cluster calculations might provide a better description of the overall
electronic structure of  insulating  3\emph{d} oxides  than different 
band structure calculations, mainly  due to a better account for correlation effects and electron-lattice coupling. 


In the 3\emph{d} oxides with the disproportionation instability (\ref{disp}) we should consider as a minimum three different many-electron configurations, or different valent states, with different types of the charge transfer. One strategy to deal with such a mixed-valence system is to create model pseudospin Hamiltonians which can reasonably well reproduce both the ground state and important low-energy excitations of the full problem. Standard pseudospin formalism represents a variant of the equivalent operators technique widely known in different physical problems from classical and quantum lattice gases, binary alloys, (anti)ferroelectrics,.. to neural networks. The formalism starts with a finite basis set for a lattice site (triplet of $M^0,M^{\pm}$ centers in our model, see below). Such an approach differs from well-known pseudospin-particle transformations akin Jordan-Wigner\,\cite{JW} or Holstein-Primakoff\,\cite{HP} transformation which establish a strict linkage between pseudospin operators and the creation/annihilation operators of the Fermi or Bose type. The pseudospin formalism for electron systems generally proceeds with a truncated basis and does not imply a strict relation to fermion operators that obey the fermionic anti-commutation rules.

To demonstrate the perspectives of the disproportionation in driving the superconductivity
we address hereafter a simplified toy  model of a mixed-valence system with three possible stable valence states of a cation-anion
 cluster (CuO$_4$, MnO$_6$, FeAs$_4$,...), hereafter  $M$: $M^0, M^{\pm}$, forming the charge (isospin) triplet and neglect all other degrees of freedom saving only the quantum charge one.  
Similarly to  the neutral-to-ionic electronic-structural transformation
  in organic charge-transfer crystals (see, e.g., paper by T. Luty in Ref.\,\cite{Toyozawa})
   the system of  charge  triplets
can  be described in frames  of the S=1 pseudo-spin formalism\,\cite{Moskvin-LTP}.
  To this end we associate three different valence charge states of the $M$-center:
  $M^0,M^{\pm}$  with three components of the $S=1$ pseudo-spin (isospin)
triplet with  $M_S =0,+1,-1$, respectively. Having in mind quasi-2D cuprates we associate 
$M^0,M^{\pm}$ centers with three charge states of the CuO$_4$ plaquette: a bare center $M^0$=CuO$_4^{6-}$, a hole center $M^{+}$=CuO$_4^{5-}$, and an electron center $M^{-}$=CuO$_4^{7-}$, respectively. 

However, the physically simple toy mixed-valence system should be described by rather complex 
effective pseudospin Hamiltonian as follows\,\cite{Moskvin-LTP}
$$
  \hat H =  \sum_{i}  (\Delta _{i}S_{iz}^2
  - h_{i}S_{iz}) + \sum_{i<j} V_{ij}S_{iz}S_{jz}+ 
$$
$$  
\sum_{i<j} [D_{ij}^{(1)}(S_{i+}S_{j-}+S_{i-}S_{j+})+ 
D_{ij}^{(2)}(T_{i+}T_{j-}+T_{i-}T_{j+})]
$$
\begin{equation}
  +\sum_{i<j} t_{ij}(S_{i+}^{2}S_{j-}^{2}+S_{i-}^{2}S_{j+}^{2}),
  \label{H}
\end{equation}
with a charge density constraint: $\frac{1}{2N}\sum _{i} \langle S_{iz}\rangle =\Delta n$, where $\Delta n$ is the deviation from a half-filling ($N_{M^+}$\,=\,$N_{M^-}$). 
Two first single-site terms describe the effects of a bare pseudo-spin splitting,
or the local energy of $M^{0,\pm}$ centers. The second term   may be
related to a   pseudo-magnetic field ${\bf h}_i$\,$\parallel$\,$Z$, in particular, a real electric field which acts as a chemical potential. 
 The third term describes the effects of the short-  and long-range inter-site density-density interactions including  screened Coulomb and covalent couplings. 
The last three kinetic energy terms in (\ref{H}) describe the one-  and two-particle  hopping, respectively ($T_{\pm}$\,=\,$\{S_z, S_{\pm}\}$).
All the parameters have a clear physical meaning. The energy $\Delta_{CT}$ for the creation of uncoupled electron and hole centers, or effective parameter $U_{dd}$ is: $\Delta_{CT}$\,=\,$2\Delta$, while the energy $\Delta_{EH}$ for the creation of coupled electron and hole centers, or EH-dimer, is: $\Delta_{EH}=2\Delta -V_{nn}$.

One should note that despite many simplifications, the
effective pseudospin Hamiltonian (\ref{H}) is rather complex, and represents one of the
most general forms of the anisotropic $S=1$ non-Heisenberg Hamiltonians.
Its real spin counterpart corresponds to an anisotropic S=1 magnet with a single ion (on-site) and two-ion (bilinear and biquadratic) anisotropy in an external magnetic field.   
Spin Hamiltonian (\ref{H}) describes an interplay of the Zeeman,   single-ion and two-ion anisotropic terms giving rise to a competition of an (anti)ferromagnetic  order along Z-axis with an in-plane $XY$ magnetic order. Simplified versions of anisotropic $S=1$ non-Heisenberg Hamiltonians have been investigated rather extensively in recent years\,\cite{Sengupta1,Sengupta2}.

 Our system is characterised by several order parameters. These are two classical ($diagonal$) order parameters: $\langle S_z
\rangle$ being a valence, or charge density with an electro-neutrality
constraint, and $\langle S_{z}^{2} \rangle$ being the density of polar centers $M^{\pm}$, or "ionicity". In addition, there are two unconventional {\it off-diagonal} order parameters 
 $\langle S_+ \rangle $ and $\langle T_+ \rangle$ related to two different types of a correlated single-particle transport, and one order parameter $\langle S_{+}^{2} \rangle$ related to a two-particle transport. It is worth noting that the ${\hat S}_{+}^{2}$ operator
creates an on-site hole pair, or composite boson, with a kinematic constraint $({\hat S}_{+}^{2})^2$\,=\,0, that underlines its "hard-core" nature. 
For real S=1 spin systems we deal with conventional "linear" magnetic order parameters $\langle S_z\rangle$ and $\langle S_+ \rangle $ related to diagonal (Z-) and off-diagonal (XY-) orderings and several spin-quadrupole order parameters:$\langle S_{z}^{2} \rangle$, $\langle T_+ \rangle$, and $\langle S_{+}^{2} \rangle$. Recently, the spin-quadrupole (spin-nematic) order $\langle S_{+}^{2} \rangle$ has been investigated in Ref.\,\cite{Sengupta2}.


The last three terms in (\ref{H}) representing the one- and two-particle hopping, respectively, are of primary importance for the transport properties, and deserve special attention. 
Two types of one-particle hopping are
governed by two transfer integrals $D^{(1,2)}$, respectively.
 The transfer integral  $t_{ij}^{\prime}=(D_{ij}^{(1)}+ D_{ij}^{(2)})$
 specifies the  probability amplitude for a {\it local disproportionation, or the $eh$-pair creation}:
$$
M^0 +M^0 \rightarrow M^{\pm}+M^{\mp}\, ,
$$
and the inverse process of the  {\it $eh$-pair recombination}:
$$
M^{\pm}+M^{\mp}\rightarrow M^0 +M^0 ,
$$
while the transfer integral  $t_{ij}^{\prime\prime}=(D_{ij}^{(1)}- D_{ij}^{(2)})$
 specifies the  probability amplitude for   a polar center transfer:
 $$
 M^{\pm} + M^0 \rightarrow M^0 +M^{\pm},
 $$
 or the {\it  motion of the electron (hole) center in the lattice of
$M^0$-centers} or motion of the $M^0$-center in the lattice of $M^{\pm}$-centers.
It should be noted that, if $t_{ij}^{\prime\prime}=0$ but $t_{ij}^{\prime}\not=0$, the eh-pair is locked in a two-site configuration.
The two-electron(hole) hopping is governed by the transfer integral
  $t_{ij}$ that defines  a probability amplitude for the exchange reaction:
$$
M^{\pm}+M^{\mp}\rightarrow M^{\mp} +M^{\pm}\, ,
$$
 or the {\it   motion of the electron (hole) center in the
lattice of the hole (electron) centers}.
It is worth noting that in the conventional Hubbard-like models
 all the types of one-electron(hole) transport are governed by the same
 transfer integral: $t_{ij}^{\prime}=t_{ij}^{\prime\prime}$, while our model implies independent parameters for the
 disproportionation/recombination process and a simple quasiparticle motion
  on the lattice of $M^0$-centers. In other words, we deal with a "correlated"   single particle transport. 


\subsection{Electron-hole Bose liquid, hard-core composite bosons and superconductivity in disproportionated systems}

Classical (or chemical)
description of the mixed-valence systems implies a full neglect of the {\it off-diagonal} purely quantum CT
effects: $D^{(1,2)}=t=0$, hence the valence of any site remains to be definite:
$0, \pm 1$, and we deal with a system of localized polar centers. Quantum description implies the taking account of  the CT effects so that we arrive at {\it
quantum superpositions of different valence states} resulting in an $indefinite$
on-site valence and ionicity whose effective, or mean values $\langle
S_{z}\rangle$ and $\langle S_{z}^2\rangle$ can vary from $-1$ to $+1$ and $0$
to $+1$, respectively.

Simple uniform 
mean-field   phases of the mixed-valent system  include an insulating monovalent $M^0$-phase (parent phase), mixed-valence binary (disproportionated) $M^{\pm}$-phase, and mixed-valence ternary ("under-disproportionated") $M^{0,\pm}$-phase\,\cite{Moskvin-LTP}.

Insulating monovalent $M^0$-phase with $ \langle S_{z}^2\rangle = 0$ is a rather conventional ground state phase for various Mott-Hubbard insulators such as 3d oxides with a large enough positive magnitude of $\Delta$\,$>$\,$\Delta_{cr}$ parameter ($U$\,$>$\,0). All the centers have the same bare $M^0$ valence state. 
Mixed-valence binary (disproportionated) $M^{\pm}$-phase with $ \langle
S_{z}^2\rangle = 1$ implies an overall disproportionation $M^0 +M^0 \rightarrow M^{\pm}+M^{\mp}$.  It is a rather unconventional
 phase for  insulators. All the centers have the  "ionized"  valence state, one half the $M^+$ state, and another half the $M^-$  one, though one may in common conceive of a deviation from the half-filling. 
 A simplified "chemical" approach to $M^{\pm}$-phase as to a classical disproportionated
phase is widely spread in solid state chemistry\,\cite{Ionov}. Typical ground state of such a classical phase corresponds to a checkerboard charge order, or longitudinal antiferromagnetic (staggered) pseudospin Ising-like Z-ordering. In systems with a short-range density-density coupling $V_{nn}$ the disproportionation phase transition $M^{0}\rightarrow M^{\pm}$, or pseudospin reorientation occurs, if $\Delta$\,$<$\,$\Delta_{cr}=\frac{z}{2}V_{nn}$, where $z$ is a number of nearest neighbours. In other words, the charge transfer instability of the parent $M^{0}$-phase implies not only negative but also small enough positive $\Delta_{CT}$\,=\,$U_{dd}$.

The mixed valence $M^{\pm}$ phase as a system of strongly correlated electron  and hole  centers
appears to be equivalent to a Bose-liquid in contrast with the electron-hole  Fermi-liquid in conventional semiconductors, hence it can be termed as {\it electron-hole Bose liquid} (EHBL)\,\cite{Moskvin-LTP,Moskvin-11,Moskvin-09}. Indeed, one may  address the electron $M^-$ center  to be a system of a local composite  boson ($e^2$) localized on the hole  $M^+$ center: $M^- = M^+ + e^2$.

Three well known  molecular-field uniform phase
states of the $M^{\pm}$ binary mixture, or EHBL phase can be specified as follows:

i) charge ordered (CO) insulating state with $\langle S_{z}\rangle = \pm
1$  and zero modulus of bosonic off-diagonal order parameter:
$|\langle S_{+}^2\rangle |= 0$;

ii) Bose-superfluid (BS) superconducting state with $\langle S_{z}
\rangle
 = 0$,  $\langle S_{+}^2\rangle = e^{2i\phi}$;

iii) mixed Bose-superfluid-charge ordering  (BS+CO) superconducting state
(supersolid)
 with $0<|\langle S_{z}\rangle |<1$, $\langle S_{+}^2\rangle \not= 0$.
 
 In addition, we should mention the high-temperature non-ordered (NO) Bose-metallic phase
 with $\ll S_z \gg =0$.
 
Disproportionation in such oxides as (Ca,Sr)FeO$_3$, RNiO$_3$ yields an electron-hole Bose system with the 50\% concentration of the composite bosons, or the half-filling. 
At variance with the parent quasi-2D cuprates the disproportionated 3d oxides from the very beginning can be addressed to be "negative-$U$", or $M^{\pm}$ systems with $\Delta$\,$<$\,$\Delta_{cr}$ where the superconductivity can be driven by a deviation from the half-filling.      
 
How a typical insulating parent 3d $M^0$-system, such as La$_2$CuO$_4$, can be driven to the disproportionated $M^{\pm}$, or EHBL phase? Simplest way is the hole/electron doping due to a nonisovalent substitution (NIS), that does solve two problems at once. First, it creates the impurity centers for a local condensation of the EH dimers, or pairs of $M^{\pm}$-centers thus shifting the system to the "negative-$U$" regime. Second, the doping in cuprates such as La$_{2-x}$Sr$_x$CuO$_{4}$ and Nd$_{2-x}$Ce$_x$CuO$_{4}$ gradually shifts the $M^{\pm}$-phase away from half-filing making the concentration of the local S-bosons to be $n_B=0.5-x/2$ (LSCO) or $n_B=0.5+x/2$ (NCCO) thus promoting the superconductivity. In both the hole- and electron-doped cuprates we deal with composite S-bosons moving on the lattice of the hole centers CuO$_4^{5-}$, that makes the unconventional properties of the hole centers\,\cite{NQR-NMR,JETPLett-12} to be common ones for the both types of cuprates. It is clear that the disproportionation scenario makes the doped cuprates the objects of  $bosonic$ physics.
Obviously, the appearance of an inhomogeneous impurity potential under  nonisovalent substitution deforms the phase diagram typical for uniform systems, in particular, in the "underdoped" \, regime where the system transforms from the parent phase $M^0$ into the disproportionated phase $M^{\pm}$.

In the limit $\Delta \rightarrow -\infty$, the EHBL phase is equivalent to the lattice hard-core (hc) Bose system with an inter-site repulsion whose Hamiltonian can be written in a standard form as follows\,(see Refs.\cite{RMP,bubble} and references therein):
\begin{equation}
\smallskip
H_{hc}=-\sum\limits_{i>j}t_{ij}{\hat
P}({\hat B}_{i}^{\dagger}{\hat B}_{j}+{\hat B}_{j}^{\dagger}{\hat B}_{i}){\hat P}
+\sum\limits_{i>j}V_{ij}N_{i}N_{j}-\mu \sum\limits_{i}N_{i},  \label{Bip}
\end{equation}
where ${\hat P}$ is the projection operator which removes double occupancy of
any site. 
Here ${\hat B}^{\dagger}({\hat B})$ are
the Pauli creation (annihilation) operators which are Bose-like commuting for
different sites $[{\hat B}_{i},{\hat B}_{j}^{\dagger}]=0,$ if $i\neq j,$  $[{\hat B}_{i},{\hat B}_{i}^{\dagger}]=1-2N_i$,
$N_i = {\hat B}_{i}^{\dagger}{\hat B}_{i}$; $N$ is a full number of sites. $\mu $  the chemical potential
determined from the condition of fixed full number of bosons $N_{l}=
\sum\limits_{i=1}^{N}\langle N_{i}\rangle $ or concentration $\;n=N_{l}/N\in
[0,1]$. The $t_{ij}$ denotes an effective transfer integral,  $V_{ij}$ is an
intersite interaction between the bosons. 
It is worth noting that near half-filling ($n\approx 1/2$) one might introduce the renormalization $N_i \rightarrow (N_i -1/2)$, or neutralizing background, that immediately provides the particle-hole symmetry.   

The model of hard-core bosons with an intersite repulsion as a minimal model of the EH Bose liquid can be mapped to a system of pseudospins $s$\,=\,1/2  exposed to an external magnetic field
in the $Z$-direction (Matsubara-Matsuda transformation\,\cite{MM}). For the system with a neutralizing background we arrive at an effective pseudo-spin Hamiltonian
\begin{equation}
H_{hc}=\sum_{i>j}J^{xy}_{ij}({\hat s}_{i}^{+}{\hat s}_{j}^{-}+{\hat s}_{j}^{+}{\hat s}_{i}^{-})+\sum\limits
_
{i>j}
J^{z}_{ij}{\hat s}_{i}^{z}{\hat s}_{j}^{z}-\mu \sum\limits_{i}{\hat s}_{i}^{z}, \label{spinBG}
\end{equation}
where $J^{xy}_{ij}=2t_{ij}$, $J^{z}_{ij}=V_{ij}$, ${\hat s}^{-}= \frac{1}{\sqrt{2}}{\hat B}_ , {\hat s}^{+}=-\frac{1}{\sqrt{2}}
 {\hat B}^{\dagger}, {\hat s}^{z}=-\frac{1}{2}+{\hat B}_{i}^{\dagger}{\hat B}_{i}$,
${\hat s}^{\pm}=\mp \frac{1}{\sqrt{2}}({\hat s}^x \pm i{\hat s}^y)$.


In terms of the $s$\,=\,1/2 pseudospins the non-ordered NO, or liquid phase, corresponds to a paramagnetic phase, Bose superfluid BS order does to a magnetic order in the $XY$ plane, while charge density order CO does to a magnetic order in the $Z$ direction. 

\begin{figure}[t]
\begin{center}
\includegraphics[width=8.5cm,angle=0]{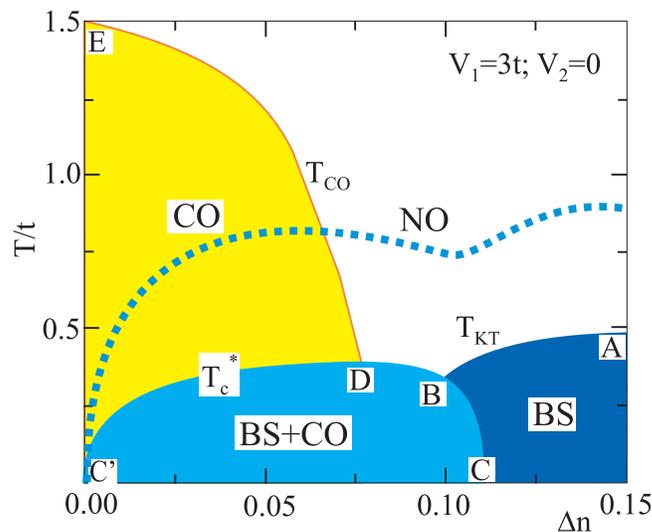}
\caption{(Color online) The QMC T-n phase diagram of the 2D hard-core boson system (reconstruction of Fig.\,2 from Ref.\cite{Schmid}, $V_1=V_{nn}, V_2=V_{nnn}$). Dotted line  schematically circumscribes the region of the well developed BS+CO and BS superconducting fluctuations (local supercurrent response) giving rise to  anomalous Nernst and local diamagnetism signals.} \label{fig1}
\end{center}
\end{figure} 
 In Fig.\,\ref{fig1} we present the phase diagram of the square lattice hc-boson Hubbard model with the nearest neighbour ($nn$) transfer integral $t_{nn}=t$ (the Josephson coupling) and repulsion $V_{nn}=3t$, derived from the quantum Monte-Carlo (QMC) calculations by Schmid {\em et al.}\,\cite{Schmid} (see, also Ref.\,\cite{RMP}). Different filling points to CO phase, BS phase, and phase separated supersolid BS+CO phase. The AB line $T_{KT}(x)$ points to 2D Kosterlitz-Thouless phase transition; the C-B-D-C$^{\prime}$ line points to the first order phase transition; the D-E line $T_{CO}(\Delta n)$  which can be termed as the pseudogap onset temperature $T^*(\Delta n)$ points to the second order Ising kind melting phase transition CO-NO into a nonordered, or normal fluid phase. 
 At half-filling ($n_B=0.5, \Delta n=0$) the system obviously prefers a checkerboard charge order (N\'{e}el antiferromagnetic order in the $Z$-direction below $T_{CO}=0.567\,V_{nn}$ given $t=0$).
 
It is worth noting that the phase diagram in Fig.\,\ref{fig1} greatly resembles that of typical for doped quasi-2D cuprates\,\cite{Honma} truly reproducing many important aspects of the normal and superconducting state, in particular, the pseudogap-like effects of the charge ordering  and  various signatures of the local superconductivity, that explain the anomalous Nernst\,\cite{ELi} and local diamagnetism signals\,\cite{diamagnetism}. 
 
\subsection{Topological phase separation in the 2D hard-core Bose system}

 It is worth noting that all the lines in the phase diagram of the 2D hard-core Bose system in Fig.\,\ref{fig1} point only to long range orders and do not concern the intricate intrinsic inhomogeneity.
However, the CO state was shown\,\cite{bubble} to be unstable with regard to a so-called CO+BS topological phase separation under doping, or deviation from the half-filling.
The boson addition or removal in the half-filled hard-core boson system  is assumed to be a driving force for a nucleation of  a  multi-center skyrmion-like self-organized
collective mode that resembles  a system of CO
bubble domains with a Bose superfluid and extra bosons both confined in domain
walls. The antiphase domain wall in the CO bubble domain appears to be a very
efficient ring-shaped potential well for the localization of a single extra
boson (or bosonic hole) thus forming a novel type of a topological defect with a neutral or
single-charged (q=2e) domain wall.
Such a  {\it topological} CO+BS {\it phase separation}, more than likely being the dynamical one, rather than an uniform mixed CO+BS supersolid phase predicted by the mean-field approximation\,\cite{RMP}, is believed to describe the evolution of hc-BH model away from half-filling. 

\begin{figure}[t]
\begin{center}
\includegraphics[width=8.5cm,angle=0]{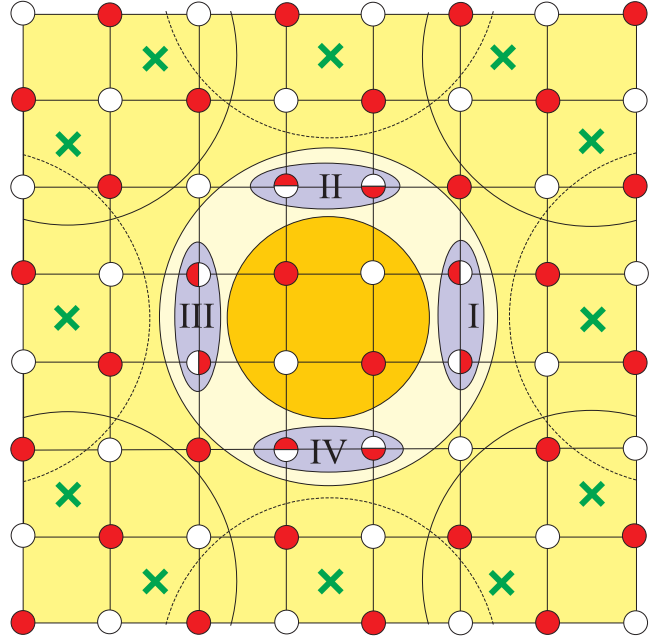}
\caption{Skyrmion-like bubble domain in the
checkerboard CO phase with the  eight site ring-shaped domain wall. Electron ($M^-$) and hole ($M^+$) centers are marked by different filling. The four dimers within the domain wall are marked by I-IV. Their specific filling points to a bond-centered CO in the domain wall rather then a site-centered CO in the domain.  Schematically shown are nearest and next-nearest neighbor domains which do not overlap with the central domain. 
Centers of all the  12 nearest neighbor domains which could be involved to a "dangerous"
frustrating overlap with the central domain are marked by crosses.}
\label{fig2}
\end{center}
\end{figure}

Fig. \ref{fig2} presents the schematic view of the smallest neutral
skyrmion-like bubble domain in a checkerboard CO phase for a 2D square
lattice with an effective size of the three lattice spacings ($\sim$\,12\,\AA)\,\cite{bubble}. The domain
wall is believed to include as a minimum eight sites forming a
ring-shaped system of four  dimers each composed of  two sites.
There are two types of the domains which differ by a rotation by
$\pm \pi /2$. Such a charged topological defect can be addressed as an extended  (q\,=\,$\pm$\,2e) skyrmion-like mobile quasiparticle. On the other hand, the bubble seems to be a peculiar quantum of a "local superconductivity".  

The bubble domain may be addressed to be well isolated only if  its close surrounding
does not contain  another domain(s) which could be involved to a "dangerous"
frustrating overlap.  Each domain in Fig.\ref{fig2} has
$z=12$ of such neighbors. Hence the concentration of well isolated domains can
be written as follows:
$
P_{0}(\Delta n)=|\Delta n|(1-|\Delta n|)^z,
$
where $z$ is a number of "dangerous" neighbors, $\Delta n$ the doped boson 
concentration. The  $P_{0}(\Delta n)$ maximum  is reached at  $\Delta n_{0}=\frac{1}{z+1}$.
In our case $\Delta n_{0}=1/13$, or $\approx 0.077$. With increasing  doping the deviation of $P_{0}(\Delta n)$
from the linear law rises. On the other hand, knowing the effective domain area
$S_d \approx 9a^2$ we can roughly estimate the limiting concentration of the
single-domain model description to be $\Delta n_{max}\approx 1/9$\,$\approx 0.11$. 

The bubble's system is believed to reveal many properties typical for granular superconductors, CDW materials,  Wigner crystals, and multi-skyrmion system akin in a quantum Hall
ferromagnetic state of a 2D electron gas\,\cite{bubble}. The concept of a static or dynamic bubble system introduces a large body of novelties into the physics of the EH Bose liquid  related  to  a complex intra-bubble structure, bubble transport,  bubble-bubble coupling, novel excitation modes, pinning of the bubble system etc. From a viewpoint of the novel charge order effects we point to an evolution of the bubble system upon lowering the temperature which implies a sequence of the isotropic liquid phase, the  liquid-crystal phase, and the  incommensurate bubble crystal phase with a quantum melting effect near the "magic" \, doping level, e.g.  $\Delta n_{b}=1/16$, where the bubble lattice undergoes the
structural phase transition. It is worth noting that the bubble crystallization is accompanied by different (pseudo)gap effects. From a viewpoint of the off-diagonal BS order we should point to a step-by-step formation of the local BS condensate starting with an  intra-EH-dimer order at $k$T\,$\sim$\,$t_{nn}$, an intra-bubble order at $k$T\,$\sim$\,$t_{nnn}$, a short-range bubble order at $k$T\,$\sim$\,$t_{bb}$ ($t_{bb}$ is a bubble transfer integral), formation of  extended quasi-1D clusters which can undergo a transition to a phase-coherent state  (filamentary or gossamer superconductivity\,\cite{Mitin-FNT}).
Finally, upon the rise of the number of such clusters, the prerequisites for a percolative 2D and bulk 3D superconductivity are created. Attractively large temperatures  of  the emergence of the nonzero local BS condensate density engender different reasonable speculations as regards its practical realization.

The topologically
inhomogeneous phase of the hc-BH system away from the half-filling
can exhibit the signatures both of $s-,d-$, and $p$\,-\,symmetry of the
off-diagonal order\,\cite{bubble}. The model allows us to study  subtle
microscopic details of the order parameter distribution including
its symmetry in a real rather than momentum space, though the
problem of the structure and stability of nanoscale domain
configurations remains to be solved. 

It is worth noting that  the bubble model predicts the optimal conditions for 2D and 3D superconductivity in the doped cuprates at $x_{opt}\approx$\,2$\Delta n_0$\,$\approx$\,0.154 and a suppression of superconductivity at $x$\,$>$\,2$\Delta n_{max}\approx$\,0.22. On the other hand, the bubble model predicts a so-called "self-doping" (SD) mechanism of superconductivity in "negative-$U$", or $M^{\pm}$ systems. Such a mechanism is realized when the CO phase in the system appears to be unstable with regard to a nucleation of positively and negatively charged bubbles preserving a charge neutrality. Probably, such a "self-doping" explains the superconductivity in YBa$_2$Cu$_4$O$_8$\,\cite{124} and LiFeAs\,\cite{LiFeAs}.

It should be noted that the bubble system can manifest two types of the dispersive features that reflect a dominance of the kinetic energy (the bubble transfer)  or the potential energy (the bubble-bubble coupling), respectively. Indeed, a single mobile bubble has energy $\varepsilon (\bf q)$, while different static bubble crystal configurations (Wigner electronic crystals) can be specified by the quasimomentum $\bf q$ and corresponding energy $E(\bf q)$.

\section{Disproportionation and local composite bosons in 3\lowercase{d}$^n$ systems}

\subsection{Local composite bosons in 3\lowercase{d}$^n$ systems}

As we have seen above the disproportionation 
$$
3d^n+3d^n \rightarrow 3d^{n+1}+3d^{n-1}
$$
is addressed to be a candidate mechanism to drive the superconductivity in nominally insulating 3d compounds. A simple view of the disproportionated system implies the electron(hole) center $3d^{n+1}$($3d^{n-1}$) to be composed of hole(electron) center plus electron(hole) coupled pair $3d^2$($3\underline{d}^2$) that seems to constitute a composite electron(hole) local  boson. In other words, the disproportionated system is anticipated to be a system of  local  bosons moving in a lattice formed by electron(hole) centers. The disproportionation scenario implies the composite local boson to be anyhow an integral part of a stable many-electron atomic configuration, $3d^{n+1}$ or $3d^{n-1}$. Microscopic nature of the "attractive force which could overcome the natural Coulomb repulsion between two electrons which constitute a Bose pair"\,\cite{Alexandrov} evolves both from main intra-atomic correlations  and electron-lattice polarisation effects\,\cite{Harrison} providing the stability of $3d^{n\pm 1}$ configurations. 

However, for the representation of local composite bosons to justify we should take into account some rather strict limitations imposed by the orbital structure of the 3d states and the genealogy of many-electron $3d^{n,n\pm 1}$ states. First of all,  the effective two-particle two-site transfer integrals
$$
\left\langle 3d_1^{n+1}3d_2^{n-1}|\hat H|3d_1^{n-1}3d_2^{n+1} \right\rangle \, ,
$$
that describe the exchange reaction 
$$
3d^{n+1}+3d^{n-1}\rightarrow 3d^{n-1}+3d^{n+1}
$$
are due to reduce to a simple form
$$
\left\langle 3d_1^{n+1}3d_2^{n-1}|\hat H|3d_1^{n-1}3d_2^{n+1} \right\rangle=
\left\langle 3d_1^{2}|\hat H|3d_2^{2} \right\rangle =t_{12}
$$
we need to introduce directly the local boson representation. The two-particle  transfer integral $t_{12}$ is a main parameter that governs both the transport and superconducting perspectives of the disproportionated system. 
It can be written as follows:
$$
t_{12}=  \langle 20| V_{ee}| 02\rangle - \sum_{11}\frac{\langle 20| \hat h| 11\rangle  \langle 11|  \hat h| 02\rangle }{\Delta_{dd}}\, ,
$$
where the first term describes a simultaneous tunnel transfer of the electron
pair due to the Coulomb coupling $V_{ee}$ and may be called as a "potential"
contribution, whereas the second describes a two-step (20-11-02) electron-pair transfer
via   successive one-electron transfer due to the one-electron Hamiltonian  $\hat
h$, and may be called as a "kinetic" contribution. The value of the composite boson transfer integral is closely related to that of the  exchange
integral, i.e. $t_{12}\approx$\,$J_{12}$.

Justification of the local composite boson   implies a specific genealogy of the many-electron states  $\Psi (3d^{n+1})$ and $\Psi (3d^{n-1})$: the wave functions should have a simple structure 
\begin{equation}
	\Psi (3d^{n+1})=\Psi (3d^{2})\Psi (3d^{n-1}), \, \Psi (3d^{n-1})=\Psi (3\underline{d}^{2})\Psi (3d^{n+1})
\label{genealogy}
\end{equation}
in electron, or hole representation, respectively. 
To gain more insight into the main features of the disproportionation reaction
hereafter we analyze the $d$\,-\,$d$ CT transition $3d^n+3d^n \rightarrow 3d^{n+1}+3d^{n-1}$
in strongly correlated 3d systems with a high (octahedral, tetrahedral, and cubic) local symmetry.
For crystal field configurations such as $t_{2g}^{n_1}e_g^{n_2}$ the optimal disproportionation scheme (\ref{genealogy}) implies  single configurations both for $\Psi (3d^{n+1})$ and $\Psi (3d^{n-1})$. It is easy to see that the optimal conditions for the superconductivity  are expected for parent $3d^n$ systems with $3d^{n\pm 1}$ configurations which correspond either empty,  or filled and half-filled $t_{2g}$ and $e_g$ orbitals. 
Second, to minimize the reduction effect of the electron-lattice interaction and avoid the localization, we need  the $S$-type ($A_{1g}$,$A_{2g}$) orbital symmetry of the local boson that provides the conservation of the orbital degeneracy when it moves on the lattice. Third, to minimize the reduction effect of the spin degrees of freedom, we need spin-singlet local bosons.  Fourth, to maximize the transfer integral we need a participation of the strongest $\sigma$ bonds. In other words, we should point to a specific role of the one-electron orbitals forming the two-electron boson state. Furthermore, the optimized ground state of the local boson should be well isolated, that is separated from excited states by a gap on the order of 1\,eV. In any case, along with the electron-lattice polarisation effects which stabilize the $3d^{n\pm 1}$ configurations and minimize the CT energy, the many-electron configuration,   intra-atomic correlations, force and symmetry of the crystal field seem to be most important factors of the disproportionation scenario. All the above constraints restrict strongly the number of 3d$^n$ systems to be candidates for the high-T$_c$ superconducting materials in frames of the disproportionation scenario.


\subsection{Optimal disproportionation schemes for different 3\lowercase{d}$^n$ ions in high-symmetry crystal field}
Below we address the optimal disproportionation schemes for different 3\lowercase{d}$^n$ ions in high-symmetry crystal field. It is worth noting that for  $octahedral$ or $tetrahedral$ ($cubic$) crystal field the electron filling starts with the $t_{2g}$ or $e_g$ orbitals, respectively. Depending on the  relation between the crystal field and intra-atomic correlation energies we will consider either high-spin (HS) or low-spin (LS) complexes such as MeO$_6$ or MeO$_4$ in 3d oxides.

{\bf 3d$^1$ ions (Ti$^{3+}$, V$^{4+}$)}.
 Disproportionation 3d$^1$+3d$^1$\,$\rightarrow$\,3d$^0$+3d$^2$ gives rise to a system of local composite bosons, or electron pairs 3d$^2$, moving in a lattice formed by the 3d$^0$ centers. Simplest electronic configuration 3d$^1$  is realized as $t_{2g}^1$ configuration for octahedral  or $e_{g}^1$ configuration for the cubic or tetrahedral crystal field. For the  $t_{2g}^2$ configuration  we arrive at an orbital and spin triplet ground state ${}^3T_{1g}$ with seemingly no optimistic expectations as regards the superconductivity. Some perspectives may be related to a spin-orbital coupling $V_{SO}$ that stabilizes a singlet state with a net momentum J=0, though the electron-lattice interaction can destroy  spin-orbital coupling giving rise to an Jahn-Teller polaron. More attractive expectations are related to the $e_g^2$ configuration in the tetrahedral crystal field when we arrive at an S-like though spin-triplet ground state ${}^3A_{2g}$. In other words, there appears an opportunity to realize an unconventional system of spin-triplet local S-like bosons moving on the electronically "inactive" lattice (see Fig.\,\ref{fig1}).  

{\bf 3d$^2$ ions (V$^{3+}$, Cr$^{4+}$)}.
Two-electron configuration 3d$^2$  is realized as $t_{2g}^2$ configuration for octahedral  or $e_{g}^2$ configuration for the cubic or tetrahedral crystal field. Disproportionation 3d$^2$+3d$^2$\,$\rightarrow$\,3d$^1$+3d$^3$ seems to gives rise to a system of electron pairs 3d$^2$, or local composite bosons, moving in a lattice formed by the 3d$^1$ centers.
However, the genealogy of the low-spin $e_g^3;{}^2E_{g}$  configuration 
does not permit the insertion of a well defined composite boson. Indeed,  the genealogical link of the $e_g^3$ and $e_g^1$ configurations reads as follows: 
$$
	\left|e_g^3;{}^2E_{g}\right\rangle =\frac{1}{\sqrt{6}}\left|(e_g^2){}^1A_{1g};e_g^1:{}^2E_{g}\right\rangle -
\frac{1}{\sqrt{3}}\left|(e_g^2){}^1E_{g};e_g^1:{}^2E_{g}\right\rangle 
$$
\begin{equation}
-\frac{1}{\sqrt{2}}\left|(e_g^2){}^3A_{2g};e_g^1:{}^2E_{g}\right\rangle \, ,
\label{e3}	
\end{equation}
and implies a participation of not  one but three different terms of the $e_g^2$  configuration.
Genealogy of the high-spin $t_{2g}^3;{}^4A_{2g}$  configuration obeys the master condition (see Exp.\,(2)) that permits the formation of a well defined composite boson with HS-configuration $t_{2g}^2;{}^3T_{1g}$. It means that the disproportionation reaction
$$
t_{2g}^2+t_{2g}^2\rightarrow t_{2g}^1+t_{2g}^3
$$
gives rise to a system of the local electron type bosons ($t_{2g}^2$) on the lattice with the on-site $t_{2g}^1$ configuration or hole type bosons ($\underline{t}_{2g}^2$) on the lattice with the on-site HS-$t_{2g}^3;{}^4A_{2g}$ configuration. However, in any case such a composite boson appears to be charged by both spin and orbital degrees of freedom, in addition, this is prone to move on the  
lattice with a spin or spin and orbital degeneracy that strongly restricts "superconducting perspectives". Some hopes may be related to a spin-orbital coupling $V_{SO}$ that stabilizes a singlet state with a net momentum J=0, though the electron-lattice interaction will compete with the  spin-orbital coupling giving rise to an Jahn-Teller polaron. 

{\bf 3d$^3$ ions (V$^{2+}$, Cr$^{3+}$, Mn$^{4+}$)}.
Three-electron configuration 3d$^3$  is realized as $t_{2g}^3$ configuration for octahedral, $e_{g}^3$ or $e_{g}^2t_{2g}$ configuration for strong or weak  tetrahedral or cubic crystal field, respectively. However, the local boson representation  can be introduced  
only for tetrahedral/cubic high-spin configuration:
$$
e_{g}^2t_{2g}+e_{g}^2t_{2g}\rightarrow e_{g}^2+e_{g}^2t_{2g}^2
$$
when the disproportionated system can be addressed to be a system of $t_{2g}^2;{}^3T_{1g}$ local composite bosons on the lattice with the on-site $e_{g}^2;{}^3A_{2g}$ 3d-configuration. Obviously, we arrive at delusive perspectives for the high-T$_c$ superconductivity.  


{\bf 3d$^4$ ions (Mn$^{3+}$, Cr$^{2+}$, Fe$^{4+}$)}.
The electron configuration near the half-filling, $3d^4$, is particularly interesting given the $octahedral$ crystal field. Then the high-spin (Hund) $3d^4$ configuration disproportionates as follows:
\begin{equation}
(t_{2g}^3e_g^1;{}^5E_g)+(t_{2g}^3e_g^1;{}^5E_g)\rightarrow
(t_{2g}^3;{}^4A_{2g})+ (t_{2g}^3e_g^2;{}^6A_{1g}) \, ,
\label{4} 
\end{equation}
where $(t_{2g}^3e_g^2;{}^6A_{1g})=(t_{2g}^3;{}^4A_{2g})\times (e_g^2);{}^3A_{2g}$.
In other words, the disproportionated system can be viewed as a system of spin-triplet composite electron bosons $(e_g^2);{}^3A_{2g}$ moving in the lattice composed of the localized hole spin-3/2 S-type $(t_{2g}^3;{}^4A_{2g})$ centers (see Fig.\,\ref{fig1}). Interestingly that we can address the system in other way to be a system of spin-triplet composite hole bosons $(\underline{e}_g^2);{}^3A_{2g}$ moving in the lattice composed of the localized electron spin-5/2 S-type $(t_{2g}^3e_g^2;{}^6A_{1g})$ centers. It is worth noting that the electron-type boson obeys a conventional (ferromagnetic) Hund rule on lattice sites while the hole-type boson obeys an unconventional (antiferromagnetic) "anti"-Hund rule on lattice sites. Local composite bosons can be formed also for the low-spin octahedral or high-spin tetrahedral $d^4$ systems, however,  we 	arrive at the composite boson with a $t_{2g}^2;{}^3T_{1g}$ or $\underline{t}_{2g}^2;{}^3T_{1g}$ configuration and a spin-3/2 or 5/2 lattice. 

{\bf 3d$^5$ ions (Fe$^{3+}$, Mn$^{2+}$)}.
The low-spin  $3d^5$ configuration appears to be near filling both in  $octahedral$ and $tetrahedral$ crystal field when it disproportionates as follows:
\begin{equation}
(t_{2g}^5)+(t_{2g}^5)\rightarrow(t_{2g}^4;{}^3T_{1g})+ (t_{2g}^6;{}^1A_{1g}) \, 
\label{5o} 
\end{equation} 
or
\begin{equation}
(e_g^4t_{2g}^1)+(e_g^4t_{2g}^1)\rightarrow(e_{g}^4;{}^1A_{1g})+ (e_g^4t_{2g}^2;{}^3T_{1g}) \, 
\label{5t} 
\end{equation} 
where $(t_{2g}^4;{}^3T_{1g})=(\underline{t}_{2g}^2;{}^3T_{1g})\times (t_{2g}^6;{}^1A_{1g})$.
In other words, the disproportionated system given the $octahedral$  crystal field can be viewed as a system of spin and orbital triplet composite hole bosons $(\underline{t}_{2g}^2;{}^3T_{1g})$ moving in a lattice composed of the spinless S-type $((t_{2g}^6;{}^1A_{1g})$ centers. Given $tetrahedral$ crystal field we arrive at a system of spin and orbital triplet composite electron bosons $(t_{2g}^2;{}^3T_{1g})$ moving in a lattice composed of the spinless S-type $((e_{g}^4;{}^1A_{1g})$ centers. Orbital degeneracy of the local bosons can give rise to the formation of the JT polarons followed by its localization. 

{\bf 3d$^6$ ions (Fe$^{2+}$, Co$^{3+}$)}.
In this  case the specific genealogy (\ref{genealogy}) of polar 3d$^{n\pm 1}$ centers can be realized only
for high-spin (Hund) configuration. 
In  $tetrahedral$ or $cubic$ crystal field the $3d^6$ system disproportionates as follows:
\begin{equation}
(e_g^3t_{2g}^3;{}^5E_g)+(e_g^3t_{2g}^3;{}^5E_g)\rightarrow
(e_g^2t_{2g}^3;{}^6A_{1g})+(e_g^4t_{2g}^3;{}^4A_{2g})  \, ,
\label{6} 
\end{equation}
where $(e_g^2t_{2g}^3;{}^6A_{1g})=(e_g^4t_{2g}^3;{}^4A_{2g})\times (\underline{e}_g^2);{}^3A_{2g}$.
In other words, the disproportionated system can be viewed as a system of spin-triplet composite hole bosons $(\underline{e}_g^2);{}^3A_{2g}$ moving in the lattice composed of the electron S=3/2 $(e_g^4t_{2g}^3;{}^4A_{2g})$ centers. 
We can address the system in other way to be a system of spin-triplet composite electron bosons $(e_g^2);{}^3A_{2g}$ moving in the lattice composed of the localized hole spin-5/2 S-type $(e_g^2t_{2g}^3;{}^6A_{1g})$ centers (see Fig.\,\ref{fig1}). The situation resembles that of the Hund $3d^4$ electron configuration given the $octahedral$ crystal field.

In octahedral crystal field the HS-$3d^6$ system disproportionates as follows:
\begin{equation}
(t_{2g}^4e_{g}^2;{}^5T_{2g})+(t_{2g}^4e_{g}^2;{}^5T_{2g})\rightarrow
(t_{2g}^3e_{g}^2;{}^6A_{1g})+(t_{2g}^5e_{g}^2;{}^4T_{1g})  \, .
\label{6a} 
\end{equation}
The disproportionated system can be viewed as a system of spin-triplet composite electron-type bosons $(t_{2g}^2;{}^3T_{1g})$ moving in the lattice composed of the hole S=5/2 $t_{2g}^3e_{g}^2;{}^6A_{1g}$ centers. Orbital degeneracy for the composite boson implies a competition of the vibronic JT coupling and the kinetic energy.

{\bf 3d$^7$ ions (Ni$^{3+}$, Co$^{2+}$)}.
For low-spin (non-Hund) configuration in  $octahedral$ crystal field the $3d^7$ system disproportionates as follows:
\begin{equation}
(t_{2g}^6e_g^1;{}^2E_g)+(t_{2g}^6e_g^1;{}^2E_g)\rightarrow
(t_{2g}^6;{}^1A_{1g})+ (t_{2g}^6e_g^2;{}^3A_{2g}) \, ,
\label{7} 
\end{equation}
where $(t_{2g}^6e_g^2;{}^3A_{2g})=(t_{2g}^6;{}^1A_{1g})\times (e_g^2);{}^3A_{2g}$.
In other words, the disproportionated system can be viewed as a system of spin-triplet composite electron bosons $(e_g^2);{}^3A_{2g}$ moving in the lattice composed of the hole S=0 $(t_{2g}^6;{}^1A_{1g})$ centers (see Fig.\,\ref{fig1}).

For high-spin (Hund) configuration in  $octahedral$ crystal field the $3d^7$ system disproportionates as follows:
\begin{equation}
(t_{2g}^5e_g^2;{}^4T_{1g})+(t_{2g}^5e_g^2;{}^4T_{1g})\rightarrow
(t_{2g}^4e_g^2;{}^5T_{2g})+ (t_{2g}^6e_g^2;{}^3A_{2g}) \, ,
\label{7a} 
\end{equation}
The disproportionated system can be viewed as a system of spin-triplet composite hole bosons $\underline{t}_{2g}^2;{}^3T_{1g}$ moving on the lattice composed of the electron S=1 $(t_{2g}^6e_g^2;{}^3A_{2g})$ centers.

{\bf 3d$^8$ ions (Ni$^{2+}$, Cu$^{3+}$)}.
For the ground state configuration in  $octahedral$ crystal field the $3d^8$ system disproportionates as follows:
\begin{equation}
(t_{2g}^6e_g^2;{}^3A_{2g})+(t_{2g}^6e_g^2;{}^3A_{2g})\rightarrow
(t_{2g}^6e_g^1;{}^2E_{g})+ (t_{2g}^6e_g^3;{}^2E_{g}) \, ,
\label{8} 
\end{equation}
with the formation of the two Jahn-Teller centers which are prone to a vibronic localization.
However, the genealogy of the $e_g^3;{}^2E_{g}$ configuration 
does not permit the insertion of the well defined composite boson (see Exp.\,\ref{e3}).
The $3d^8$ system given the $tetra hedral$ crystal field disproportionates as follows:
\begin{equation}
(e_{g}^4t_{2g}^4;{}^3T_{1g})+(e_{g}^4t_{2g}^4;{}^3T_{1g})\rightarrow
(e_{g}^4t_{2g}^3;{}^4A_{2g})+ (e_{g}^4t_{2g}^5;{}^2T_{2g}) \, ,
\label{8a} 
\end{equation}
In other words, such a disproportionated system can be viewed as a system of spin-triplet composite electron-type JT bosons $(t_{2g}^2;{}^3T_{1g})$ moving on the lattice composed of the hole spin-3/2   $(e_{g}^4t_{2g}^3;{}^4A_{2g})$ S-type centers.
\begin{figure}[t]
\begin{center}
\includegraphics[width=15cm,angle=0]{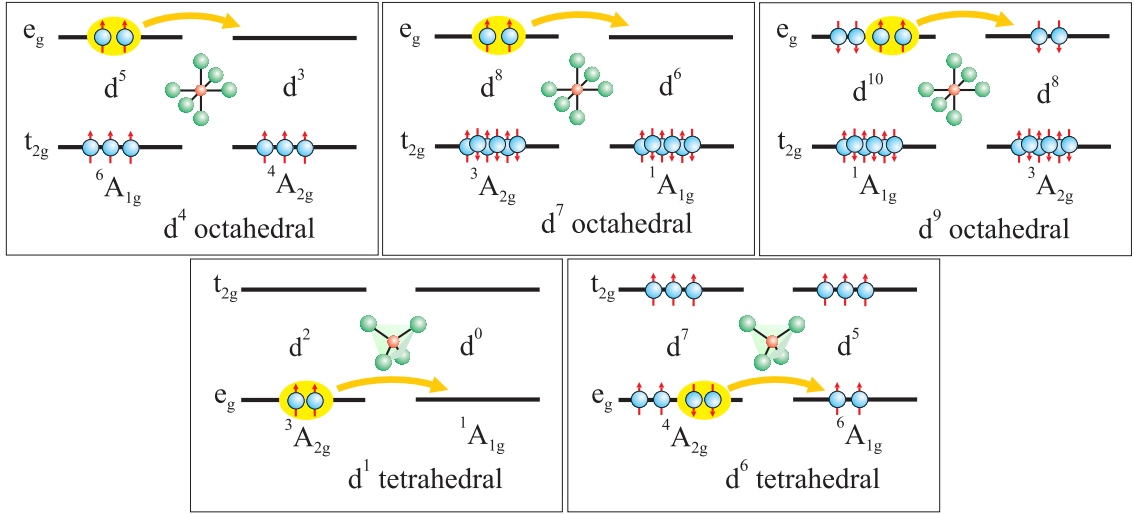}
\caption{(Color online) Disproportionation schemes for the high-symmetry crystal field which are believed to provide optimal conditions for the disproportionation driven superconductivity.} \label{fig3}
\end{center}
\end{figure}
 
{\bf 3d$^9$ ions (Cu$^{2+}$)}.
In  $octahedral$ crystal field the $3d^9$ system disproportionates as follows:
\begin{equation}
(t_{2g}^6e_g^3;{}^2E_g)+(t_{2g}^6e_g^3;{}^2E_g)\rightarrow
(t_{2g}^6e_g^2;{}^3A_{2g})+ (t_{2g}^6e_g^4;{}^1EA_{1g}) \, ,
\label{9octa} 
\end{equation}
where $(t_{2g}^6e_g^2;{}^3A_{2g})=(t_{2g}^6e_g^4;{}^1A_{1g})\times (\underline{e}_g^2);{}^3A_{2g}$.
In other words, the disproportionated system can be viewed as a system of spin-triplet composite hole bosons $(e_g^2);{}^3A_{2g}$ moving in the lattice composed of the electron spin-0  $(t_{2g}^6e_g^4;{}^1A_{1g})$ centers (see Fig.\,\ref{fig1}).

In  $tetrahedral$ or $cubic$ crystal field the $3d^9$ system disproportionates as follows:
\begin{equation}
(e_g^4t_{2g}^5;{}^2T_{2g})+(e_g^4t_{2g}^5;{}^2T_{2g})\rightarrow
(e_g^4t_{2g}^4;{}^3T_{1g})+ (e_g^4t_{2g}^6;{}^1A_{1g}) \, ,
\label{9tetra} 
\end{equation}
where $(e_g^4t_{2g}^4;{}^3T_{1g})=(e_g^4t_{2g}^6;{}^1A_{1g})\times (\underline{t}_{2g}^2);{}^3T_{1g}$.
In other words, the disproportionated system can be viewed as a system of spin-triplet composite hole bosons $(\underline{t}_{2g}^2);{}^3T_{1g}$ moving in a lattice composed of the electron spinless  $(e_g^4t_{2g}^6;{}^1A_{1g})$ centers.

\subsection{Summary}

Let's summarize our findings.

i) The most effective $e_g^2$ configuration of the local composite boson providing maximal values of the boson transfer integral without strong reduction effects of the electron-lattice coupling is realized only for several $optimal$ $d^n$ configurations. These are HS-$d^4$, LS-$d^7$, $d^9$ configurations given octahedral crystal field, and $d^1$, HS-$d^6$ configurations given tetrahedral crystal field\,\cite{Moskvin-FPS-08};

ii) All these bare systems are characterized by E-type orbital degeneracy, i.e. these are prone to a strong Jahn-Teller effect. In all the instances  the  disproportionation reaction lifts the bare orbital degeneracy, that is it has a peculiar "anti-Jahn-Teller" character\,\cite{Mazin}.

iii) All these bare systems are characterized by a strong suppression due to a vibronic reduction of the one-particle (electron or hole) transport, while after disproportionation these are characterized by an effective two-particle (local boson) transport.

iv) Undesirable spin-triplet structure of the local composite bosons appears to be a common feature of all the above mentioned disproportionated systems with a high-symmetry crystal field.

v) At variance with $d^1$, $d^7$, and $d^9$ systems the $d^4$ and $d^6$ systems reveal unavoidable coexistence of the spin-triplet bosons and a magnetic lattice.

All these results are summarized in Table\,I where we have added examples of 3d systems with optimal $3d^n$ configurations. 
The Fig.\,\ref{fig3} illustrates the final electronic configurations and two-particle $e_g^2$ transport for the disproportionation reaction in optimal $d^{4,6,7,9}$ systems.  
\begin{center}
\begin{table}
\caption{$3d^n$ JT-systems optimal for the disproportionation driven superconductivity (see text for detail)}
\begin{tabular}{|c|c|c|c|c|c|c|}
\hline
   \begin{tabular}{c}
Electron \\
configuration \\  
\end{tabular}        & Symm. & LS/HS & \begin{tabular}{c}
Local \\
boson \\  
\end{tabular}  & Lattice& \begin{tabular}{c}
Parent(bare) \\
compounds \\  
\end{tabular}   & 
SC    \\ \hline
  \begin{tabular}{c}
$3d^1$($e_g^1$):${}^2E$ \\
$Ti^{3+}$, $V^{4+}$ \\  
\end{tabular}  & tetra& -&\begin{tabular}{c}
$e_g^2$:${}^3A_{2g}$ \\
s=1 \\  
\end{tabular}  & S=0 & ? & ?\\ \hline
  \begin{tabular}{c}
$3d^4$($t_{2g}^3e_g^1$):${}^5E$ \\
$Mn^{3+}$, $Fe^{4+}$ \\  
\end{tabular}& octa& HS&\begin{tabular}{c}
$e_g^2$:${}^3A_{2g}$ \\
s=1 \\  
\end{tabular}   & S=3/2 & \begin{tabular}{c}
(Ca,Sr)FeO$_3$\\
RMnO$_3$ \\  
\end{tabular} & \begin{tabular}{c}
NIS, SD \\
T$_c$, (?) \\  
\end{tabular}  \\ \hline
  \begin{tabular}{c}
$3d^6$($e_g^3t_{2g}^3$):${}^5E$ \\
$Fe^{2+}$, $Co^{3+}$ \\  
\end{tabular}& tetra& HS&\begin{tabular}{c}
$e_g^2$:${}^3A_{2g}$ \\
s=1 \\  
\end{tabular}   & S=5/2 & \begin{tabular}{c}
LaFeAsO,...\\
LiFeAs \\  
\end{tabular}    & \begin{tabular}{c}
NIS, SD \\
T$_c$$\leq$\,56\,K \\  
\end{tabular}  \\ \hline
  \begin{tabular}{c}
$3d^7$($t_{2g}^6e_g^1$):${}^2E$ \\
$Co^{II+}$, $Ni^{III+}$ \\  
\end{tabular}& octa& LS&\begin{tabular}{c}
$e_g^2$:${}^3A_{2g}$ \\
s=1 \\  
\end{tabular}  & S=0 & \begin{tabular}{c}
RNiO$_3$\\
AgNiO$_2$ \\  
\end{tabular} & ?\\ \hline
  \begin{tabular}{c}
$3d^9$($t_{2g}^6e_g^3$):${}^2E$ \\
$Cu^{2+}$ \\  
\end{tabular}& octa$^*$& -&\begin{tabular}{c}
$\underline{b}_{1g}^2$:${}^1A_{1g}$ \\
s=0 \\  
\end{tabular}  & S=0 & \begin{tabular}{c}
La$_2$CuO$_4$,...\\
YBa$_2$Cu$_4$O$_8$ \\  
\end{tabular} & \begin{tabular}{c}
NIS, SD \\
T$_c$$\leq$\,135\,K \\  
\end{tabular}    \\ \hline 
       \end{tabular}
\end{table}
\end{center}

Our consideration does support and generalize a purely electronic motivation for "negative $U$"\, centers which was based upon a stabilization of the electron-rich closed-shell configurations such as $ns^2$ or $nd^{10}$ through the charge excitation effects or of half-filled configurations such as $nd^5$ through the exchange-correlation effects (Hund's rule)\,\cite{negative-U}. The approach does explain the missing oxidation states of ions in $s^1$ electronic configuration and predict the disproportionational, or charge transfer  instability in systems with ions having nominally $ns^1$ (Hg$^+$, Tl$^{2+}$, Pb$^{3+}$, Bi$^{4+}$) , $d^9$ (Cu$^{2+}$), $d^4$ (Mn$^{3+}$, Fe$^{4+}$), $d^6$ (Fe$^{2+}$, Co$^{3+}$) valence electronic configuration.

In the above we proceed with a conventional approach to the electronic structure of the 3d compounds that implies a  predominantly 3d antibonding character of the molecular $t_{2g}$ and $e_g$ orbitals in octahedral MO$_6$ or tetrahedral MO$_4$ clusters. Electron and spin density in these clusters is distributed between the 3d ion and ligands that should be taken into account, in particular, when address a so-called "partial" disproportionation such as $Ni^{(3+\delta )+}$\,-\,$Ni^{(3-\delta )+}$ in RNiO$_3$\,\cite{Alonso}. 
However, upon lowering the crystal field symmetry and anomalous strengthening of the  $dp$\,-hybridization in the end of the $d^n$ series the situation can change in several important points. Remarkable example is provided by quasi-2D cuprates where we proceed with a strong axial distortion of the CuO$_6$ octahedra, actually with square CuO$_4$ plaquettes ($D_{4h}$ point symmetry) as a basic element of the crystal and electronic structure. Single hole $b_{1g}(\propto d_{x^2-y^2})$-state of Cu$^{2+}$ ion is a typical one for the square-planar coordination of oxygen ions whereas two-hole Cu$^{3+}$ state with the same coordination of oxygen ions seldom exists. 
Instability of Cu$^{3+}$ ions with 3d$^8$, or two-hole 3$\underline{d}^2$ configuration is obviously related to a strong inter-electron(hole) repulsion. However, the two-hole state in CuO$_4^{5-}$ center to be a cluster analogue of Cu$^{3+}$ ion can be stabilized by a location of the additional hole to low-energy predominantly oxygen O 2p molecular orbitals  thus providing a sharp suppression of the inter-electron(hole) repulsion at a relatively small loss in the one-hole energy.  
In 1988 Zhang and Rice\,\cite{ZR} have proposed that the doped hole forms a well isolated local spin and orbital ${}^1A_{1g}$ singlet state which involves a phase coherent combination of the 2p$\sigma$ orbitals of the four nearest neighbor oxygens with the same $b_{1g}$ symmetry as for a bare Cu 3$d_{x^2-y^2}$ hole.  
The spin and orbital  singlet  $\underline{b}_{1g}^2;{}^1A_{1g}$, or Zhang-Rice singlet, may be viewed as a local hole composite boson. In other words, the disproportionation $3d^9+3d^9\rightarrow 3d^{10}+3d^{8}$ in the tetragonally distorted octahedral $3d^9=3\underline{d}^{1}$ (Cu$^{2+}$) systems gives rise to a system of composite hole S=0 bosons whose spin and orbital structure provides   maximal values of the effective  boson transfer integral and minimal boson effective mass, that is optimal conditions for the disproportionation driven high-T$_c$ superconductivity.
Such a situation is unlikely realized for the $d^1$ configuration  in a tetrahedral crystal field because of a weak \emph{p-d} covalency.  
 Thus, our analysis shows that in fact only $d^9$  system can be a major candidate for maximal T$_c$'s.
 
Above we did not concern the energy stability of the final disproportionated "negative-$U$" \, state. It should be noted that together with a sizeable recombination energy this implies well isolated ground states of $d^{n\pm 1}$ configurations.
 
\section{Electron-hole Bose liquid with spin-triplet $s$\,=\,1 local composite bosons}

Above we have presented a unified approach to the disproportionation phenomenon in different 3d compounds though we did not touch upon such important points as electron and lattice polarization effects, vibronic coupling, and the role played by the spin and orbital degrees of freedom. All these can strongly deform the phase diagram of the 3d compounds from the predictions of the purely charge order parameter model.

Minimal model of the EHBL phase does not imply intervention of the spin degrees of freedom only in quasi-2D cuprates. Indeed, the model considers such a cuprate to be a system of the the spin and orbital singlet ${}^1A_{1g}$   local S-bosons moving on the lattice formed by hole centers with the well isolated spin and orbital singlet Zhang-Rice ${}^1A_{1g}$ ground state. For all other "optimal" 3d$^n$ systems listed in Table I we arrive at an unavoidable spin-triplet $s$\,=\,1 structure of the composite local bosons moving on the spin (bare octa-HS d$^4$ and tetra-HS d$^6$ configurations) or spinless (bare tetra- d$^1$ and octa-LS d$^7$ configurations) lattice.  
In the absence of the external
magnetic field the effective Hamiltonian of such an  electron-hole Bose liquid  takes the form of the Hamiltonian of the quantum lattice Bose gas of the triplet bosons with an exchange coupling\,\cite{Moskvin-09}:
$$ \hat H = \hat H_{QLBG}+\hat
H_{ex}=\sum_{i\not=j,m}t_{B}(ij)\hat B_{im}^{\dagger}\hat B_{jm}
$$
$$
+\sum_{i>j}V_{ij}n_{i}n_{j}
-\mu\sum_{i}n_{i} 
+\sum_{i>j}J_{ij}^{hh}(\hat{\bf S}_{i}\cdot\hat{\bf S}_{j}) 
$$
\begin{equation}
+
\sum_{i\not=j}J_{ij}^{hb}(\hat{\bf s}_{i}\cdot\hat{\bf S}_{j})
+\sum_{i>j}J_{ij}^{bb}(\hat{\bf s}_{i}\cdot\hat{\bf s}_{j})
+\sum_{i}J_{ii}^{hb}(\hat{\bf s}_{i}\cdot\hat{\bf S}_{i})\,.
\label{HH}
\end{equation}
Here $\hat B_{im}^{\dagger}$ denotes the $s$\,=\,1 boson creation operator with a spin projection $m$  at the site $i$; $\hat B_{im}$ is the corresponding annihilation operator. The boson number operator $\hat n_{im}=\hat B_{im}^{\dagger}\hat B_{im}$ at $i$-site
due to the condition of the on-site infinitely large repulsion  $V_{ii}\rightarrow+\infty$ ({\it hardcore boson}) can take values 0 or 1. 

The first term in (\ref{HH}) corresponds to the kinetic energy of the bosons,
$t_B(ij)$ is the transfer integral. The second one reflects the effective
repulsion ($V_{ij}>0$) of the bosons located on the neighboring sites. The
chemical potential $\mu$ is introduced to fix the boson concentration:
$
n=\frac{1}{N}\sum_{i}\langle \hat n_{i}\rangle\,.
$
For EHBL phase in parent system we arrive at the same number of electron and hole centers, that is to $n=\frac{1}{2}$.
The  remaining terms in  (\ref{HH}) represent the Heisenberg exchange
interaction between the spins of the hole centers (term with $J^{hh}$), spins
of the hole centers and the neighbor boson spins (term with $J^{hb}$), boson
spins (term with $J^{bb}$), and the very last term in (\ref{HH}) stands for the
intra-center Hund exchange between the boson spin and the spin of the hole center.
In order to account for the Hund rule one should consider $J^{hb}_{ii}$ to be
infinitely large ferromagnetic. 

Generally speaking, this model Hamiltonian describes the system that can be
considered as a Bose-analogue of the {\it one orbital} double-exchange model system\,\cite{Dagotto}.

More or less simple manifestation of spin degrees of freedom are anticipated for tetra-$d^1$ and LS-octa-$d^7$ configurations where we proceed with spin-triplet composite bosons, however, moving on the spinless lattice.

Estimates for different superexchange couplings given the bond geometry typical for manganites such as LaMnO$_3$\,\cite{Moskvin-09} predict antiferromagnetic coupling of the $nn$ hole centers ($J^{hh}>0$), antiferromagnetic coupling of the two nearest neighbor bosons  ($J^{bb}>0$), and ferromagnetic coupling of the boson and the nearest neighbor hole centers ($J^{hb}<0$).
In other words, we arrive at highly frustrated system of triplet bosons moving in a lattice formed by hole centers when the hole centers  tend to order G-type antiferromagnetically, the triplet bosons tend to order ferromagnetically both with respect to its own site and its nearest neighbors. Furthermore, nearest neighboring bosons strongly prefer an antiferromagnetic ordering. Lastly, the boson transport prefers an overall ferromagnetic ordering.

Competition of the Heisenberg exchange and the bosonic double exchange in "optimal" systems such as ferrates (Ca,Sr)FeO$_3$ or manganites RMnO$_3$ with bare octa-HS d$^4$ configurations of the transition metals can be easily demonstrated on an example of the $d^5$\,-\,$d^3$ pair, or EH-dimer.
The net spin of the EH-dimer is ${\bf S}={\bf S }_1+{\bf S}_2$, where ${\bf S }_1$ ($S_1=5/2$) and ${\bf S }_2$ ($S_2=3/2$) are spins of Fe$^{3+}$ and Fe$^{5+}$ (Mn$^{2+}$ and Mn$^{4+}$) ions, respectively. In nonrelativistic approximation the spin structure of the EH-dimer will be determined by isotropic Heisenberg exchange coupling 
\begin{equation}
V_{ex}=J\,({\bf S }_1\cdot {\bf S }_2),	
\end{equation}
 with $J$ being an exchange integral, and the two-particle charge transfer characterized by a respective transfer integral which depends on spin states as follows:
\begin{equation}
\langle \frac{5}{2}\frac{3}{2};SM|\hat H_B|\frac{3}{2}\frac{5}{2};SM\rangle =\frac{1}{20}S(S+1)\,t_B\, ,	
\end{equation}
 where  $t_B$ is a spinless  transfer integral. Making use of this relation we can introduce an effective spin-operator form for the boson transfer as follows:
 \begin{equation}
	\hat H_B^{eff}=\frac{t_B}{20}\left[2(\hat {\bf S}_1\cdot\hat {\bf S}_2)+S_1(S_1+1)+S_2(S_2+1)\right]\, ,
\end{equation}
 which can be a very instructive tool both for qualitative and quantitative analysis of boson transfer effects, in particular, the temperature effects.  
 Both conventional Heisenberg exchange coupling and unconventional two-particle bosonic transfer, or bosonic double exchange can be easily diagonalized in the net spin S representation so that for the energy we arrive at
\begin{equation}
E_S=\frac{J}{2}[S(S+1)-\frac{25}{2}]\pm \frac{1}{20}S(S+1)\,t_B\,,	
\end{equation}
where $\pm$ corresponds to two quantum superpositions $|\pm\rangle $ written in a spin representation as follows
\begin{equation}
|SM\rangle _{\pm} =\frac{1}{\sqrt{2}}(|\frac{5}{2}\frac{3}{2};SM\rangle \pm |\frac{3}{2}\frac{5}{2};SM\rangle )\, ,	
\end{equation}
with $s$- and $p$-type symmetry, respectively.  It is worth  noting that the bosonic double exchange contribution formally corresponds to a ferromagnetic exchange coupling with $J_B=-\frac{1}{10}|t_B|$. 
 \begin{figure}[t]
 \begin{center}
\includegraphics[width=12cm,angle=0]{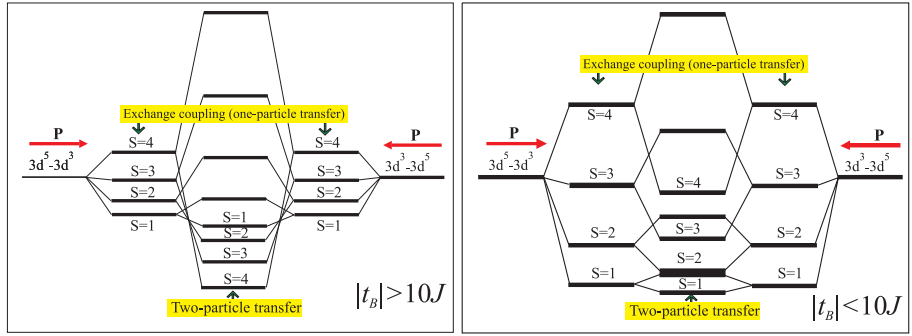}
\caption{(Color online) Spin structure of the EH-dimer, or self-trapped CT exciton with a step-by-step  inclusion of one- and two-particle charge transfer. Arrows point to electric dipole moment for bare site-centered dimer configurations.} \label{fig4}
\end{center}
\end{figure}

We see that the cumulative effect of the Heisenberg exchange and the bosonic double exchange results in a stabilization of the S\,=\,4 high-spin
 (ferromagnetic) state of the EH-dimer provided $|t_B|>10J$ and  the S\,=\,1 low-spin (ferrimagnetic) state otherwise (see Fig.\,\ref{fig4}). As in conventional (one-electron) double exchange model\,\cite{de Gennes}  the latter frustration  between the Heisenberg exchange and the bosonic double exchange stabilizes a noncollinear ordering in bulk systems.
 
\section{Discussion} 

Our approach provides a comprehensive understanding of the well established disproportionation in ferrates (Ca,Sr)FeO$_3$ and nickellates RNiO$_3$ with octa-HS $d^4$ and octa-LS $d^7$ configurations of the transition metals. It is worth noting that the flexible perovskite structure of both compounds facilitates a screening of the effective correlation parameter $U_{dd}$. In (Ca,Sr)FeO$_3$ the effect is believed to strengthen due to a large enough polarizability of the Ca$^{2+}$ or Sr$^{2+}$ ions while in RNiO$_3$ we deal with a sizeable reduction of the effective $U_{dd}$ due to a strong Ni-O covalency.

An incommensurate helicoidal spin ordering observed both in CaFeO$_3$ and SrFeO$_3$ up to very low temperatures  can be explained  as a result of a competition between conventional exchange coupling and the bosonic double exchange in EH dimers $M^{\pm}$\,$\leftrightarrow$\,$M^{\mp}$. 
Interestingly, the helicoidal spin ordering gives rise to  a full (SrFeO$_3$) or partial (CaFeO$_3$) suppression of the charge order resulting in a metallic or insulating temperature dependence of the resistivity, respectively\,\cite{CaSrFeO3}. In any case both theoretical and experimental study of the phase diagram for (Ca,Sr)FeO$_3$ system deserves further work, especially, aimed to a search of a possible superconductivity.

Nominal electronic HS-octa-configuration $d^4$ of the Fe$^{4+}$ ion is analogous to that of Mn$^{3+}$ ion in perovskite (ortho)manganites RMnO$_3$ that implies similar disproportionation features. Indeed, similarly to ferrates (Ca,Sr)FeO$_3$ the orthomanganites RMnO$_3$ reveal  a high-temperature metallic  phase which may be associated with a fully disproportionated R(Mn$^{2+}$Mn$^{4+})$O$_3$ phase\,\cite{Good,Moskvin-09}, or EHBL system. Such a conclusion is particularly supported by an anomalously small magnitude of the thermopower. Interestingly that under spin ordering this manganite phase in contrast with the ferrates could become a {\it ferromagnetic EHBL metal} as bosonic double exchange in   manganese EH dimers overcomes the Heisenberg exchange in favour of the ferromagnetic spin state\,\cite{Moskvin-09}.
In other words, in manganites we deal with the high-spin ferromagnetic ground state of the EH dimer (see Fig.\,\ref{fig4}a), while in ferrates with the low-spin ferrimagnetic one (see Fig.\,\ref{fig4}b). However, actually upon lowering the temperature one observes a first order phase transition at T=T$_{JT}$ (T$_{JT}$\,$\approx$\,750\,K in LaMnO$_3$) from the high-temperature  fully disproportionated 
metallic EHBL phase to a low-temperature orbitally ordered  insulating   phase with a cooperative Jahn-Teller ordering of the occupied $e_g$-orbitals of the Mn$^{3+}$O$_6$ octahedra accompanied by A-type antiferromagnetic
ordering below T$_N$ (T$_{N}$\,$\approx$\,140\,K in LaMnO$_3$)\,\cite{Good,Moskvin-09}. In other words, the vibronic (Jahn-Teller) coupling in bare (parent) Mn$^{3+}$ phase suppresses the anticipated charge order and does promote a recombination $M^{\pm}\rightarrow M^0$ transition.

Interestingly that the nonisovalent substitution and/or nonstoichiometry seems to revive the disproportionated  phase and such manganites along with a metallic ferromagnetism with a colossal magnetoresistance reveal many properties typical for superconducting materials. These are a steep jump of the specific heat,   gap opening in tunneling, and  giant proximity effect. Several manganite samples were shown to exhibit a negative diamagnetic  susceptibility in a wide temperature range up to room temperature\,\cite{Markovich}. To explain many experimental manifestations of superconductivity Kim \cite{Kim} has proposed a superconducting condensation in manganites.  In this scenario, colossal magnetoresistance in La$_{1-x}$Sr$_x$MnO$_3$ is naturally explained by the superconducting
fluctuations with increasing magnetic fields. This idea is closely related to the observation of anomalous proximity effect between superconducting YBCO and a  manganese oxide, La$_{1-x}$Ca$_x$MnO$_3$  or La$_{1-x}$Sr$_x$MnO$_3$\,\cite{Kasai}, and also the concept of a local superconductivity manifested by doped manganites\,\cite{Mitin}.

A strong evidence of the revived EHBL phase below first order phase transition $M^{\pm}\rightarrow M^0$ from the high-temperature EHBL phase to low-temperature AFM insulating phase is obtained very recently by Nath {\it et al.}\,\cite{Muk}. The authors observed an electric field driven destabilization of the insulating state in nominally pure LaMnO$_3$ single crystal with a moderate field which leads to a resistive state transition below 300\,K. The effect has been explained as bias driven percolation type transition between two coexisting phases, where the majority phase is a charge and orbitally ordered polaronic AFM insulating phase and the minority phase is a bad metallic EHBL phase\,\cite{Muk}.

Very recently distinct signatures of high-temperature disproportionated EHBL phases are revealed also in other manganites, such as  LaMn$_7$O$_{12}$\,\cite{LaMn7O12} with quadruple perovskite 
structure and YBaMn$_2$O$_6$\,\cite{YBaMn2O6}.
A dramatic suppression of the Curie-Weiss susceptibility in the undistorted phase of LaMn$_7$O$_{12}$ above T$_{JT}$\,$\approx$\,650\,K is explained by a picture of long-lived low-spin S=1 EH dimers (see Fig.\,\ref{fig4}b) formed by the charge transfer between neighboring Mn sites\,\cite{LaMn7O12}. 
High-temperature electron spin resonance studies of A-site ordered mixed-valent manganite YBaMn$_2$O$_6$\,\cite{YBaMn2O6} have revealed two puzzles. First, in the
metallic high-temperature phase the ESR signal stems only from all Mn$^{4+}$ core spins. Second, the ESR linewidth obeys the  Korringa-like linear high-temperature increase 
over a wide temperature range of 400\,K  clearly pointing to a truly metallic state. Both findings strongly support our suggestion of the high-temperature EHBL phase of manganites to be a system of spin-triplet composite $e_g^2$ bosons moving on the lattice of spin-3/2 Mn$^{4+}$ centers. 
Furthermore, one may speculate that the Korringa-like behavior of the ESR linewidth could be a
signature of such a charge-disproportionated picture\,\cite{YBaMn2O6}.
Our prediction of the  disproportionation driven high-T$_c$ superconductivity in the HS-$d^4$ octa-systems needs in a more direct validation for the hole doped manganites such as La$_{1-x}$Sr$_x$MnO$_3$.

High-temperature metallic phase of rare earth nickelates RNiO$_3$ with LS-octa-$d^7$ configuration of Ni$^{III+}$ ions resembles that of orthomanganites, however, with a Pauli-like magnetic susceptibility and  evidences a full \emph{d-d} disproportionation. Yet the low-temperature behavior of manganites and nickellates strongly differs. 
For rhombohedral LaNiO$_3$  the system remains metallic down to 1.5\,K,
whereas  orthorhombic RNiO$_3$ (R =Pr,... Lu) exhibit a first order phase transition to a charge ordered  insulating state upon cooling below T$_{CO}$ spanning from 130\,K for Pr to $\sim$\,550-600\,K for heavy rare earths.
Both the lighter and heavier orthorhombic RNiO$_3$ compounds are very similar from the point of view of their local electronic and magnetic CO state despite the strong change of the metal-to-insulator NO-CO transition temperature. All these exhibit clear signatures of the charge disproportionated state with two types of Ni centers 
corresponding to alternating large NiO$_6^{10-}$ (Ni$^{2+}$ center) and small NiO$_6^{8-}$ (Ni$^{4+}$ center) octahedra strongly differing in magnetic moments ($\sim$\,2\,$\mu_B$ and  $\sim$\,0, respectively). Regarding the magnetic order, these nickellates adopt an
antiferromagnetic structure defined by the propagation vector
(1/2,0,1/2) at a transition temperature T$_N$ spanning from 130\,K   to \,$\sim$\,200\,K. Antiferromagnetic ordering can be explained by a rather strong superexchange $nnn$ (next-nearest neighbor) coupling of magnetic S=1 Ni$^{2+}$ centers. Unfortunately, the antiferromagnetism stabilizes the charge order and suppresses the superconductivity.

At variance with the charge transfer unstable perovskites ((Ca,Sr)FeO$_3$, RMnO$_3$, RNiO$_3$)  the quasi-2D nickellates ANiO$_2$ (A\,=\,Ag, Li, Na) reveal  existence of unconventional ground states stabilized by the frustrated triangular lattice geometry  from a cooperative JT 
ordering   of Ni$^{3+}$ ions in NaNiO$_2$ to a {\it moderately charge ordering} 3Ni$^{III+}$\,$\rightarrow$\,Ni$^{2+}$+2\,Ni$^{3.5+}$ in antiferromagnetic metal AgNiO$_2$\,\cite{AgNiO2}. In the case of LiNiO$_2$ there could be a competition between
charge ordering and orbital ordering for the ground state, the nickel valency could be a mixture of 2+, 3+, and 4+\,\cite{LiNiO2}.

In all the  CT-unstable 3d compounds with $optimal$ electron configurations addressed above the JT coupling in nearly octahedral complexes (FeO$_6$, MnO$_6$, NiO$_6$) competed with the charge disproportionation and in several compounds this determined the orbitally ordered ground state and the low-temperature phase thus suppressing the anticipated superconductivity.  
The JT coupling reaches its maximal magnitude in Cu$^{2+}$ octahedral complexes that seemingly 
 should inhibit the disproportionation in Cu$^{2+}$ based cuprates. However, the giant JT distortion stabilizes the CuO$_4$ plaquette with a $b_{1g}\propto d_{x^2-y^2}$ hole ground state which can be a proper starting orbital for an in-plane hole transfer $\underline{b}_{1g}$\,$\rightarrow$\,$\underline{b}_{1g}$ in the CuO$_2$ planes of the quasi-2D cuprates. As a result we arrive at a disproportionation 2\,CuO$_4^{6-}$\,$\rightarrow$\,CuO$_4^{7-}$+CuO$_4^{5-}$ with formation of a spin and orbital singlet electron center CuO$_4^{7-}$ (analog of the Cu$^+$ ion) and a hole center CuO$_4^{5-}$ (analog of the Cu$^{3+}$ ion) in a Zhang-Rice ${}^1A_{1g}$
ground state. Full disproportionation within a CuO$_2$ plane results in a system of  composite two-electron spin-singlet (S-type) bosons moving on the spin-0 lattice formed by hole CuO$_4^{5-}$ centers. We arrive at a EHBL system which is the most favorable one for the high-T$_c$ superconductivity. 
A comprehensive argumentation of the disproportionation scenario for quasi-2D cuprates is reported in Ref.\,\cite{Moskvin-11}.

At first sight the situation similar to cuprates  can be realized  in a hypotetical quasi-2D nickellate with a system of well isolated Ni$^{3+}$O$_2$ planes provided a low-spin $t_{2g}^6e_g$ configuration of the NiO$_4^{5-}$ center  with the $b_{1g}\propto d_{x^2-y^2}$ electron ground state and a spin-singlet ZR-type configuration of the $e_g^2$-subshell of the electron NiO$_4^{6-}$ center. However, the latter condition seems unlikely to be realized in practice therefore the perspectives of the cuprate-like superconductivity in nickellates look like very illusive\,\cite{Ni-HTSC}.

In the above we addressed "optimal" systems with bare octahedral $d^{4,7,9}$ complexes. What about predicted  "optimal" $d^1$ and $d^6$ systems with tetrahedral local symmetry? At present we have no literature data about  such $d^1$ systems whereas the systems with tetrahedral $d^6$, or more precisely, high-spin Fe$^{2+}$ complexes number tens of the so-called iron pnictides ("FePn", where Pn is As or P) and iron chalcogenides ("FeCh", where Ch includes S, Se, and
Te)\,\cite{Fe}. Actually all these compounds are either superconductors or parent systems which turn into superconducting phase with T$_c$ values up to 56\,K under nonisovalent doping and external or  "chemical" \, pressure.

All the families of iron-containing superconductors have two-dimensional planes of FePn$_4$ or FeCh$_4$ tetrahedra. The quasi-2D structure makes the iron compounds in common with superconducting cuprates, however, the FePn/Ch superconductors are fundamentally different  from the cuprates in several points. First of all it concerns the coexistence of superconductivity with magnetism demonstrated at an unprecedented atomic scale. 

Unlike the case of the high-T$_c$ cuprates, electrons in FePn/Ch layers are always itinerant and there is no Mott insulating state in the electronic phase diagram though most of the FePn/Ch systems start with an AFM order as the parent phase. The magnetic order in FePn/Ch layers is unconventional one, it is most likely a spin-density wave with strongly reduced mean values of a local effective spin.

All the main features of the superconducting and normal state for the FePn/Ch compounds agree with our predictions of the disproportionation within the FePn/Ch layers accompanied by formation of EHBL system with a topological phase separation, e.g. formation of a multi-bubble system. However, in contrast with the spinless scenario for cuprates in the FePn/Ch layers both composite electron $e_g^2$ bosons and the hole lattice have nonzero spins, s\,=\,1 and S\,=\,5/2, respectively. It means that the charge ordered bubbles jointly are magnetic bubbles with a complex nanoscopically inhomogeneous spin structure. Accordingly, the FePn/Ch layers can be considered as  a system of the  bubble carriers which carry both charge, "local superconductivity", and a spin density. Moreover, the bubble formation is accompanied by strong  nanoscale structural distortions which can be detected by pair distribution function analysis\,\cite{Egami}.



In any case, finding the high-T$_c$ superconductivity in the FePn/Ch compounds with the $tetrahedral$ coordination of the iron Fe$^{2+}$($3d^6$) ions in HS state and its $coexistence$ with an unconventional magnetism can be a key argument supporting the disproportionation scenario\,\cite{remark}. Puzzlingly, our model implies the superconducting carriers in  FePn/Ch compounds are composed of the $e_g$ rather than $t_{2g}$ electrons as predicted by one-electron band models. Both cuprates and ferro-pnictides/chalcogenides are believed to be  "bad"\, EHBL  metals.

\section{Conclusion}
We analyzed $d$\,-\,$d$ disproportionation phenomena in materials with different 3d ions.
We have shown that optimal conditions for the  disproportionation driven high-T$_c$ superconductivity are realized  only for several $d^n$ configurations. These are HS-$d^4$, LS-$d^7$, $d^9$ configurations given octahedral crystal field, and $d^1$, HS-$d^6$ configurations given tetrahedral crystal field. Interestingly, all these bare systems are characterized by E-type orbital degeneracy, i.e. these are prone to a strong Jahn-Teller effect. In all the instances  the  disproportionation reaction lifts the bare orbital degeneracy, that is it has a peculiar "anti-Jahn-Teller" character. All the optimal systems are characterized by a strong suppression of the one-particle transport, while after disproportionation these are characterized by an effective two-particle transport.
Anyhow the optimal disproportionation schemes (see Fig.\,\ref{fig3} and Table I) imply a participation of a spin degree of freedom. Proposed superconductivity in the $d^4$ and $d^6$ systems at variance with $d^1$, $d^7$, and $d^9$ systems  is characterized by unavoidable coexistence of the spin-triplet composite bosons and a magnetic lattice.
Only quasi-2D cuprates in a "negative-$U$" regime seem to represent a system of spin-singlet composite bosons moving on the spinless lattice. Their magnetic response could be explained by purely oxygen orbital magnetism if we take into account deviations from the simple Zhang-Rice model\,\cite{NQR-NMR,JETPLett-12}. Interestingly, our "spinless" scenario for cuprates  clearly opposes to a spin/magnetic mechanism to be currently a leading contender for the superconducting mechanism in the high T$_c$ cuprates.

We argue that unconventional superconductivity observed in iron-based layered pnictides and chalcogenides with tetrahedrally coordinated Fe$^{2+}$ ions can be a key argument supporting the disproportionation scenario is at work in these compounds. It is worth noting that the potential for high-T$_c$ appears to be the greatest for undistorted FePn$_4$ tetrahedra\,\cite{Johnston}.
 Furthermore, observation of the high-T$_c$ superconductivity  in these quasi-2D systems with undesirable spin degrees of freedom both for the composite boson and the lattice gives some hope to finding of superconductivity in artificial 2D superlattices of all the "optimal" configurations.

We did not concern here many important points of the disproportionation scenario and some comparisons with other model approaches, particularly with the bipolaronic theory by Alexandrov\,\cite{Alexandrov}. 
Nevertheless the model approach suggested is believed to provide a conceptual framework for an in-depth understanding on equal footing  of physics of very different strongly correlated 3d systems such as cuprates, manganites, nickellates, ferro-pnictides/chalcogenides, and other systems with a charge transfer  instability and/or mixed valence. In particular, our paper provides a clear answer to the question, "what is so special about Cu in cuprates or Fe in the Fe-based superconductors?" and why, e.g., BaMn$_2$As$_2$ which with respect to many properties is situated between BaFe$_2$As$_2$ and the high-T$_c$ cuprates is not a high-T$_c$ superconductor\,\cite{Johnston}. 

\ack
The  RFBR grants Nos. 10-02-96032 and 12-02-01039  are acknowledged for providing financial support.


\section*{References}
\begin{thebibliography}{99}

\bibitem{Muller}  
Bednorz J G and M\"{u}ller K A 1986 
  {\it Z. Phys.} B {\bf 4} 189


\bibitem{Kamihara} 
Kamihara Y Watanabe T Hirano M  and Hosono H 2008 {\it J. Am. Chem. Soc.} {\bf 130} 3296


\bibitem{Moskvin}
Moskvin A S 1998 {\it Physica} B {\bf 252} 186;
Moskvin A S and Ovchinnikov A S 1998 {\it \JMMM} {\bf 186} 288; {\it Physica} C {\bf 296} 250; Moskvin A S and Panov Yu D 1999 {\it \PSS}
(b) {\bf 212} 141;{\it J. Phys. Chem. Solids} {\bf 60} 607



\bibitem{Moskvin-LTP}
Moskvin A S 2007 {\it Low Temp. Phys.} {\bf 33} 234


\bibitem{Moskvin-11}
Moskvin A S 2011 {\it \PR} B {\bf 84} 075116

\bibitem{Ogg} 
Ogg Jr R A 1946 {\it \PR} {\bf 69} 243

\bibitem{Schafroth} 
Schafroth M R 1955  {\it \PR} {\bf 100} 463

\bibitem{negative-U}
Hirsch J E and Scalapino D J 1985 {\it \PR} B {\bf 32} 5639;
Varma C M 1988 {\it \PRL} {\bf 61} 2731;
Khomskii D 1997 {\it Lithuanian Journal of Physics} {\bf 37} 65;
Wilson John A 2000 {\it \JPCM} {\bf 12} R517;
Geballe T H and Moyzhes B Y 2000 {\it Physica} C {\bf 341-348} 1821;
Larsson S 2002 {\it Int. J. Quantum Chem.} {\bf 90} 1457;
{\it Physica} C {\bf 460-462} 1063;
Mitsen K V and Ivanenko O M 2004 {\it Phys. Usp.} {\bf 47} 493;
Tsendin K D Popov B P and Denisov D V 2006 {\it Supercond. Sci. Technol} {\bf 19} 313;
Katayama-Yoshida Hiroshi Kusakabe Koichi Kizaki Hidetoshi and Nakanishi Akitaka 2008 {\it Appl. Phys. Express} {\bf 1} 081703



\bibitem{dd-CT}
Moskvin A S M\'{a}lek J Knupfer M Neudert R Fink J  Hayn R Drechsler S-L Motoyama N  Eisaki H  and Uchida S 2003 {\it \PRL} {\bf 91} 037001; Kovaleva N N  Boris A V Bernhard C Kulakov A Pimenov A Balbashov A M Khaliullin G and Keimer B 2004 {\it \PRL} {\bf 93} 147204; Sokolov V I Pustovarov V A Churmanov V N Ivanov V Yu Gruzdev N B Sokolov P S Baranov A N and Moskvin A S 2012 {\it JETP Letters} {\bf 95} 528;2012 {\it \PR} B {\bf 86} 115128 



\bibitem{Shluger}
Shluger A L  and Stoneham A M 1993 {\it \JPCM} {\bf 5} 3049



\bibitem{Mizokawa}
Mizokawa T Khomskii D I and Sawatzky G A 2000 {\it \PR} B {\bf 61}  11263
 
\bibitem{Alonso}
Alonso J A Mart\'{i}nez-Lope M J Casais M T Garc\'{i}a-Mu$\tilde{n}$oz J L Fern\'{a}ndez-D\'{i}az M T  and  Aranda M A G 2001 {\it \PR} B {\bf 64} 094102;Zhou J-S Goodenough J B and Dabrowski B 2003 {\it \PR} B {\bf 67} R020404
 

\bibitem{Ionov}
Ionov S P Ionova G V Lubimov V S and Makarov E F 1975 {\it \PSS} (b) {\bf 71} 11


\bibitem{CaFeO3}
Takano M Nakanishi N Takeda Y Naka S and Takada T 1977 {\it Mat Res. Bull.} {\bf 12} 923


\bibitem{Sr3Fe2O7}
Kuzushita K Morimoto S  Nasu S and Nakamura S 2000 {\it \JPSJ} {\bf 69} 2767
 
\bibitem{thermopower}
Granados X Fontcuberta J Obradors X Manosa Ll and Torrance J B 1993 {\it \PR} B {\bf 48} 11666

\bibitem{DyNiO3}
Mu$\tilde{n}$oz A Alonso J A Mart\'{i}nez-Lope M J and
 Fern\'{a}ndez-D\'{i}az M T 2009 {\it  J. Solid State Chem.} {\bf 182} 1982



\bibitem{Good}
Zhou J-S and Goodenough J B 1999 {\it \PR} B {\bf 60} R15002;  {\it \PR} B {\bf 68} 144406




\bibitem{Moskvin-09}
Moskvin A S 2009 {\it \PR} B {\bf 79} 115102

\bibitem{BaBiO3}
Hase I and  Yanagisawa T 2007  {\it \PR} B {\bf 76} 174103

\bibitem{MgO}
Stoneham A M and Sangster M J L 1981 {\it Philos. Mag.} B {\bf 43} 609

 \bibitem{Harrison}
Harrison W A 2006 {\it \PR} B {\bf 74} 245128

\bibitem{Ballhausen}
Ballhausen C J 1962 {\it Introduction to ligand field theory}, McGraw-Hill, p.298

\bibitem{JW}
Jordan P and Wigner E 1928 {\it Z. Physik} {\bf 47} 631 

\bibitem{HP}
Holstein T and Primakoff H 1940 {\it \PR} {\bf 58} 1098

\bibitem{Toyozawa}
Luty T 1997 {\it Relaxations of Excited States and Photo-Induced Structural Phase
Transitions}
ed.: K. Nasu, Springer Series in Solid-State Sciences {\bf 124} 17

\bibitem{Sengupta1}
Sengupta P and Batista C D 2007 {\it \PRL} {\bf 98} 227201; 2008 {\it \JAP} {\bf 103}  07C709

\bibitem{Sengupta2}
Wierschem K Kato Y Nishida Y  Batista C D  Sengupta P 2012 {\it arXiv:1209.0688v2}

\bibitem{NQR-NMR}
Moskvin A S 2004 {\it JETP Lett.} {80} 697
  
\bibitem{JETPLett-12}
Moskvin A S 2012 {\it JETP Lett.} {96} 385

\bibitem{RMP}  
Micnas R Ranninger J and Robaszkiewicz S 1990 {\it \RMP} {\bf 62} 113
 
\bibitem{bubble}
Moskvin A S Bostrem I G and Ovchinnikov A S 2003 {\it JETP Lett.} {\bf 78} 772; Moskvin A S 2004 {\it \PR} B {\bf 69} 214505

\bibitem{MM}
Matsubara T and Matsuda H 1956 {\it Prog. Theor. Phys.} {\bf 16} 569

\bibitem{Schmid}
Schmid G Todo S Troyer M and Dorneich A 2002 {\it \PRL} {\bf 88} 167208
    
    

\bibitem{Honma}
Honma T H Hor P 2008 {\it \PR} B {\bf 77} 184520

\bibitem{ELi}
Li E  Sharma R P  Ogale S B  Zhao Y G  Venkatesan T  Li J J
Cao W L and Lee C H 2002 {\it \PR} B {\bf 65} 184519;  Li E  Sharma R P  Ogale S B   Krivoruchko V N and  Petryuk R V 2002 {\it \PR} B {\bf 66} 134520
  
\bibitem{diamagnetism}
Li L Wang Y Komiya S Ono S Ando Y Gu G D and Ong N P 2010 {\it \PR} B {\bf 81} 054510

\bibitem{Mitin-FNT}
Mitin A V 2007 {\it Low Temp. Phys.} {\bf 33} 245

\bibitem{124}
Karpinski J Kaldis E Jilek E Rusiecki S and Bucher B 1988 {\it Nature (London)} {\bf 336} 660


\bibitem{LiFeAs}
Tapp J H Tang Z Lv Bing Sasmal K Lorenz B Chu P C W and Guloy A M 2008 {\it \PR} B {\bf 78} 060505

\bibitem{Alexandrov}
Alexandrov A S 2011 {\it \PS} {\bf 83} 038301
  
    
\bibitem{Moskvin-FPS-08} 
Moskvin A S Avvakumov I L 2008 {\it Proceedings of III International conference "Fundamental problems of high-temperature superconductivity", 13-17 October 2008, Moscow-Zvenigorod} 215



\bibitem{Mazin} 
Mazin I I Khomskii D I Lengsdorf R Alonso J A Marshall W G Ibberson R M Podlesnyak A Mart\'{i}nez-Lope M J and Abd-Elmeguid M M 2007 {\it \PRL} {98} 176406

\bibitem{ZR}
Zhang F C and  Rice T M 1988 {\it \PR} B {\bf 37} 3759

\bibitem{Dagotto}
Dagotto E Hotta T and Moreo A 2001 {\it \PR} {\bf 344} 1 

\bibitem{de Gennes}
de Gennes P G 1960 {\it \PR} {\bf 118} 141

\bibitem{CaSrFeO3}
Fujioka J Ishiwata S Kaneko Y Taguchi Y and Tokura Y 2012 {\it \PR} B {\bf 85} 155141

\bibitem{Markovich}
Markovich V Fita I Wisniewski A Puzniak R Mogilyansky D Titelman L Vradman L Herskowitz M and Gorodetsky G 2008 {\it \PR} B {\bf 77} 014423

\bibitem{Kim}
Kim Yong-Jihn 1998 {\it Modern Physics Letters} B {\bf 12} 507

\bibitem{Kasai}
Kasai M Ohno T Kauke Y Kozono Y Hanazono M and Sugita Y 1990 {\it Jpn. J. Appl. Phys.} {\bf 29} L2219

\bibitem{Mitin}
Mitin A V Kuz'micheva G M and Novikova S I 1997 {\it Russian Journal of Inorganic Chemistry} {\bf 42} 1791

\bibitem{Muk}
Nath R Raychaudhuri A K   Mukovskii Ya M  Mondal P  Bhattacharya D
and  Mandal P   {\it arXiv:1212.1001v1}

\bibitem{LaMn7O12}
Cabassi R Bolzoni F Gilioli E Bissoli F Prodi A and Gauzzi A 2010 {\it \PR} B {\bf 81} 214412

\bibitem{YBaMn2O6}
Schaile S Krug von Nidda H-A Deisenhofer J Loidl A Nakajima T and Ueda Y 2012 {\it \PR} B {\bf 85} 205121

\bibitem{AgNiO2}
Wawrzynska E  Coldea R Wheeler E M Mazin I I Johannes M D Sorgel T Jansen M Ibberson R M and Radaelli P G 2001 {\it \PRL} {99} 157204

\bibitem{LiNiO2}
Chen H  Freeman C L and Harding J H 2011 {\it \PR} B {\bf 84} 085108


\bibitem{Ni-HTSC}
Chaloupka J and Khaliullin G 2008 {\it \PRL} {100} 016404


\bibitem{Fe}
Stewart G R 2011 {\it \RMP} {\bf 83} 1589

\bibitem{Egami}
Niedziela J L   McGuire M A and  Egami T 2012 {\it arXiv:1211.5179v1}



\bibitem{remark} 
The suggestion of the disproportionation driven superconductivity in iron-pnictides was first made in 2008, see Ref.\,\cite{Moskvin-FPS-08}


\bibitem{Johnston}
Johnston D C 2010 {\it Advances in Physics} {\bf  59} 803


\end{thebibliography}
\end{document}